\documentclass[fleqn,usenatbib]{mnras}

\usepackage{newtxtext,newtxmath}

\usepackage[T1]{fontenc}

\DeclareRobustCommand{\VAN}[3]{#2}
\let\VANthebibliography\thebibliography
\def\thebibliography{\DeclareRobustCommand{\VAN}[3]{##3}\VANthebibliography}

\usepackage{graphicx}	
\usepackage{amsmath}	
\usepackage{caption}
\usepackage{subcaption}
\usepackage{multicol}
\usepackage{lipsum}  
\usepackage{pdflscape}
\usepackage{xcolor}
\usepackage{soul}

\usepackage{array}
\newcolumntype{L}{>{\centering\arraybackslash}m{0.4cm}}
\newcolumntype{n}{>{\centering\arraybackslash}m{1cm}}

\title[MeerKAT follow-up of enigmatic GLEAM 4-Jy (G4Jy) sources]{MeerKAT follow-up of enigmatic GLEAM 4-Jy (G4Jy) sources}

\author[P. K. Sejake et al.]{Precious K. Sejake$^{1}$\thanks{E-mail: sejakekatlego@gmail.com},
Sarah V. White$^{1}$,
Ian Heywood$^{1,2,3}$,
Kshitij Thorat$^{4}$,
Hertzog L. Bester$^{1,2}$,
\newauthor Sphesihle Makhathini$^{1,2,5}$,
and Bernie Fanaroff$^{2}$
\\
$^{1}$Department of Physics and Electronics, Rhodes University, PO Box 94, Grahamstown, 6140, South Africa\\
$^{2}$South African Radio Astronomy Observatory (SARAO), 2 Fir Street, Observatory, Cape Town, 7925, South Africa\\
$^{3}$Astrophysics, University of Oxford, Denys Wilkinson Building, Keble Road, Oxford, OX1 3RH\\
$^{4}$Department of Physics, University of Pretoria, Hatfield, Pretoria, 0028, South Africa\\
$^{5}$School of Physics, University of the Witwatersrand, Johannesburg, Braamfontein, 2000, South Africa
}

\date{Accepted 2022 November 16. Received 2022 November 16; in original form 2022 August 20}

\pubyear{2023}

\begin{document}
\label{firstpage}
\pagerange{\pageref{firstpage}--\pageref{lastpage}}
\maketitle

\begin{abstract}
We present the results from studying 140 radio sources in the GLEAM (GaLactic and Extragalactic All-sky MWA [Murchison Widefield Array]) 4-Jy (G4Jy) Sample. These sources were followed-up with MeerKAT to assess their radio morphology and enable host-galaxy identification, as existing radio images of 25 to 45-arcsec resolution do not provide sufficient information. We refer to these sources as the MeerKAT-2019 subset. The aim is to identify the host galaxy of these sources by visually inspecting the overlays comprising radio data from four surveys (at 150, 200, 843/1400, and 1300 MHz). Our morphological classification and host-galaxy identification relies upon the $\sim$7-arcsec resolution images from MeerKAT (1300 MHz). Through the visual inspection of the overlays, 14 radio sources in the MeerKAT-2019 subset have wide-angle tail (WAT) morphology, 10 are head-tail, and 5 have X-, S-/Z-shaped morphology. Most of the remaining sources have the radio morphology of typical symmetric lobes. Of 140 sources, we find host galaxies for 98 sources, leaving 42 with no identified host galaxy. These 42 sources still have ambiguous identification even with higher resolution images from MeerKAT. 

\end{abstract}

\begin{keywords}
galaxies: formation -- galaxies: evolution -- galaxies: active -- galaxies: jets
\end{keywords}




\section{Introduction}
Two main galaxy populations are known to radiate vast amounts of radio emission: star-forming galaxies and galaxies with an active galactic nucleus (AGN), with AGNs being the dominant emitter. Both populations play a tremendous role in galaxy evolution. The impact of star formation on galaxy evolution is through the collapse of the molecular clouds to form stars, thereby building stellar mass. However, stellar mass growth may be halted (usually in galaxies with low mass) if, for instance, the supernova is powerful enough to expel gas from the system \citep{2000MNRAS.317..697E} or the gas is stripped away due to the interaction with another galaxy \citep{1991ApJ...377...72M}. \\

\noindent
Meanwhile, the process of material falling onto the supermassive black hole (SMBH) at the galaxy's centre may sometimes produce relativistic radio jets \citep{1995PASP..107..803U, 2015ARAandA..53..365N}. AGN activity has been shown to significantly impact the host galaxy by suppressing or promoting star formation; however, this is still a point of contention. \citet{2013Sci...341.1082M} demonstrated that star formation might be suppressed whereby the radio jets expel gas from the system. Meanwhile, \citet{Croft_2006} showed that the jets might trigger the collapse of the molecular gas and promote star formation. \\

\noindent
In order to gain a better acumen of which of these scenarios is more significant, a large sample of AGN with powerful relativistic radio jets is needed to provide robust statistics on the features of these AGN, taking into account attributes such as jet inclination with respect to the plane of the galaxy \citep{2014AandA...567A.125G}, the size of the molecular gas reservoirs \citep{Emonts_2011}, and jet power \citep{2016MNRAS.461..967M}. Additionally, we need a large sample to unravel the effects of jet power, age, and environmental density on radio luminosity and investigate these properties as a function of redshift. Moreover, the mechanism of how jets are launched is still poorly understood, as well as how these sources evolve with redshift. This is because most studies are constrained to the 173 'radio loud' AGN in the revised Third Cambridge Catalogue of Radio Sources (3CRR, $\text{S}_{178~\text{MHz}}>10.9~\text{Jy}$; \citealt{1983MNRAS.204..151L}). Therefore, this work is significant in that it will help to better populate the high-luminosity end.\\

\noindent
AGN activity occurs in at least two modes, leading to two intrinsically different AGN populations. This dichotomy is most likely caused by the difference in the efficiency of the mechanism governing accretion onto the central SMBH (e.g. see review by \citealt{2014ARAandA..52..589H}). The first AGN activity mode is `quasar mode', also called `cold mode', `radiative mode' or `high-excitation'. The material is accreted onto the central SMBH through a radiatively efficient, optically thick and geometrically thin disc \citep{1973AandA....24..337S}. These radiatively efficient AGN are typically connected to the most luminous AGN, radiating across a broad range of the electromagnetic spectrum. A fraction of these AGN are radio-loud AGN with powerful radio jets extending for tens or hundreds of kpc. The second AGN activity mode is `radio mode', also called `jet mode' or `low-excitation'. In this mode, the accretion of material onto the central SMBH produces little radiated energy but can result in the production of highly energetic jets. These radiatively inefficient AGN are connected to low to intermediate luminosity AGN. The evolution of the low-luminosity AGN has been probed by several studies using deeper, narrow-area surveys such as VLA-COSMOS 3GHz Large project \citep{2017AandA...602A...6S} and LoTSS Deep Field \citep{2019AandA...622A..17S, 2022MNRAS.513.3742K}.\\

\noindent
\cite{2020PASA...37...17W, 2020PASA...37...18W} have created a large sample of AGNs with the purpose of identifying how the properties of such sources vary as a function of redshift and/or environment. This sample is known as the GLEAM (GaLactic and Extragalactic All-sky MWA [Murchison Widefield Array; \citealt{2013PASA...30....7T}]; \citealt{2017MNRAS.464.1146H}) 4-Jy (G4Jy) Sample and is ten times larger than the 3CRR, due to its lower flux density limit and larger survey area ($24,731\,\text{deg}^2$). The 3CRR sample, selected at low radio frequency and complete with optical data, has made it possible to conduct groundbreaking research such as the link between radio jet power and optical luminosity \citep{1991Natur.349..138R}, which indicates that extragalactic radio sources share a central engine mechanism responsible for their emission. Moreover, the 3CRR sample was used in the study of \cite{1989ApJ...336..606B} to show that a unification model based on the orientation of the AGN can explain the observed properties of quasars and radio galaxies. \\

\noindent
The G4Jy Sample, which is a complete sample above 4-Jy at 151 MHz, is made up of 1,863 of the brightest radio sources selected from the GLEAM survey using the extragalactic catalogue (Galactic latitude, |b| > $10^{\circ}$). The GLEAM survey uses observations from MWA, the precursor telescope for the low-frequency component of the SKA (Square Kilometre Array). MWA surveyed the entire southern sky (Declination, Dec < $30^{\circ}$) at low radio frequencies in the range 72 - 231 MHz. The resolution of the GLEAM survey is declination dependant and is approximated by $2.5\times 2.2\,\text{arcmin}^2/\text{cos}(\delta+26.7^{\circ})$ at the central frequency of 154 MHz, which corresponds to a synthesised beam of $\sim$2-arcmin at 200 MHz. The G4Jy Sample does not suffer from orientation bias as these sources are selected at low frequencies. Lower frequency observations do not suffer from Doppler-boosting effects, a phenomenon inherent in surveys at higher frequencies. \\

\noindent
\cite{2020PASA...37...17W, 2020PASA...37...18W} performed the host-galaxy identification for 1,606 radio sources in the G4Jy Sample through visual inspection of overlays comprising a set of radio contours overlaid onto mid-infrared WISE (Wide-field Infrared Survey Explorer; \citealt{Wright_2010}) W1 band images. The sets of radio contours used are from; GLEAM ($\sim$2-arcmin resolution), NVSS (NRAO VLA Sky Survey [45-arcsec resolution]; \citealt{Condon_1998}) or SUMSS (Sydney University Molonglo Sky Survey [45-arcsec resolution]; \citealt{2003MNRAS.342.1117M,2007MNRAS.382..382M}) and TGSS ADR1 (TIFR GMRT Sky Survey first alternative data release [25-arcsec resolution]; \citealt{2017AandA...598A..78I}). The host galaxy identification of the remaining sources was constrained by the poor resolution provided by NVSS/SUMSS and TGSS. Moreover, there were some discrepancies in the literature concerning the host galaxy for some of the sources. As a result, higher-resolution observations are needed to perform host galaxy identification of these sources. The main objective of this study is to identify the host galaxies of a subset of 140 G4Jy radio sources, which we regard as the MeerKAT-2019 subset. The MeerKAT-2019 subset sources showed complex `enigmatic' radio morphologies at 45 and 25-arcsec resolution images. These sources were observed with MeerKAT (1.3 GHz and $\sim7-\text{arcsec}$ resolution), providing better sensitivity and angular resolution than TGSS and NVSS/SUMSS.


\subsection{Paper outline}
This paper is outlined as follows. Section \ref{sec:data} gives a brief overview of the radio and infrared data we used to construct the overlays. The MeerKAT observations and data reduction are briefly described. Section \ref{sec:method} explains how the MeerKAT-2019 subset was constructed, the overlays creation, and our labels for host-galaxy identification. The results, of which we give comments on the most interesting sources, are given in section \ref{sec:results}. The discussion and conclusion are given in sections \ref{sec:discussion} and \ref{sec:conclussion}, respectively. Throughout this paper, a flat Lambda-CDM model was assumed: $\mathrm{H_0} = 70~\mathrm{km/s/Mpc}$, $\Omega_m = 0.3$ and $\Omega_{\Lambda} = 0.7$ \citep{2016AandA...594A..13P}, and we use the sign convention $S\propto \nu^\alpha$, where $\alpha$ is the spectral index and $S$ is the integrated flux density at frequency $\nu$.

\section{Data} \label{sec:data}
In this section, we briefly describe (1) the MeerKAT observations and data reduction and (2) additionally the data sets used to perform the host galaxy identification of our sample.

\subsection{Radio data}\label{subsec:radio_data}
\subsubsection{MeerKAT observations (1.3 GHz)}
\label{subsubsec:meerkat}
We observed\footnote{Project code: SCI-20190418-SW-01, PI: White} 140 objects from the G4Jy Sample using the L-band (856 -- 1712 MHz) receivers of MeerKAT \citep{jonas2016}. A combination of the brightness of the G4Jy sources, and the high instantaneous sensitivity and excellent ($u$,$v$) plane coverage of MeerKAT meant that 5-minute snapshot observations per target were sufficient to achieve the two primary goals of unambiguous host galaxy identification, and morphological classification. Targets were observed in four blocks, grouped by Right Ascension. The overhead encompassed scans of the primary calibrator (either PKS B1934$-$638 or PKS 0408$-$65, depending on the RA), an RA-dependent secondary calibrator that was typically observed after every five target scans, the primary polarisation calibrator (either 3C 138 or 3C 286), and slewing time. The telescope correlator was configured to deliver 8-second integrations and 4,096 frequency channels, with the latter being averaged by a factor of 4 prior to processing. \\

\noindent
The data were processed in standard fashion, with the raw MeerKAT visibilities for each of the four blocks being converted to Measurement Set format using the KAT Data Access Library\footnote{\url{https://github.com/ska-sa/katdal}}. Standard flagging commands were applied to all targets to remove the low gain edges of the telescope's bandpass response, as well as to flag known regions of persistent radio frequency interference (RFI) on interferometer spacings below 600~m. Autoflagging software was then used on the calibrator sources, after which the calibrators were used to derive instrumental delay, gain, and bandpass corrections. Application of these corrections to the calibrators themselves was followed by another round of autoflagging on the residual (corrected$-$model) visibilities, after which the instrumental corrections were re-derived and applied to the targets. The calibrated target data were then flagged and imaged. A multifrequency clean component model derived from deconvolution within a masked region was used to perform phase and delay self-calibration, and the corrected data were re-imaged and finally primary beam corrected. A Briggs robust value \citep{briggs1995} of $-$0.3 was used, except for a small number of targets where uniform (robust~=~$-$2.0) weighting was used to increase angular resolution and reduce residual side lobe emission that remained due to calibration deficiencies. \\

\noindent
The self-calibrated data were also re-imaged in eight frequency chunks with data on spacings below 164 wavelengths discarded from the inner region of the ($u,v$) plane. The resulting cube was used to produce a spectral-index image for each source. This was achieved by masking all pixels where the Stokes I brightness was below 1 mJy beam$^{-1}$, and subsequently extracting the spectrum of each sight-line through the cube and fitting for its slope in log-brightness vs log-frequency space.\\

\noindent
Flagging was performed using the {\sc casa} \citep{mcmullin2007} and {\sc tricolour}\footnote{\url{https://github.com/ska-sa/tricolour/}} packages. Referenced calibration made use of the {\sc casa} package, with the {\sc cubical} \citep{kenyon2018} software used for self-calibration. All imaging made use of the {\sc wsclean} \citep{offringa2014} software. Primary beam correction was done in the image plane, using an azimuthally averaged image of the Stokes I primary beam, evaluated at the nominal band centre frequency of 1284~MHz using the {\sc eidos} \citep{asad2021} package. Our data processing scripts are available online\footnote{\url{https://github.com/IanHeywood/oxkat}} \citep{heywood2020}, and can be consulted for further details of the reduction process. These scripts were also used to deploy the data processing jobs on the two high-performance computing facilities used for this work, namely the ilifu\footnote{\url{http://www.ilifu.ac.za}} cloud computing facility, and the Centre for High-Performance Computing (CHPC\footnote{\url{https://www.chpc.ac.za/}}) \emph{Lengau} cluster.

\subsubsection{GLEAM catalogue and images (72-231 MHz)}
The GLEAM survey constitutes observations from MWA, which observed the entire southern sky Dec. < $30^{\circ}$ at low radio frequencies in the range of 72-231 MHz. We use 200 MHz images with a resolution of $\sim$2 arcmin of the GLEAM survey for visual inspection in this work. GLEAM survey covers 24,831 square degrees for Dec. < $30^{\circ}$ and |b| > $10^{\circ}$. The survey is 99.97\% reliable above a $5\sigma$ ($\sim$50 mJy) threshold, 90\% complete at 170 mJy, and 50\% complete at 55 mJy.

\subsubsection{TGSS ADR1 catalogue and images (150 MHz)}
TGSS is made up of observations from the Giant Metrewave Radio Telescope (GMRT; \citealt{1991ASPC...19..376S}), which surveyed the sky above Dec. = $-55^{\circ}$. Observations at Dec. > $-53^{\circ}$ were retained for the ADR1 due to the poor data quality at low elevations. Moreover, TGSS has incomplete coverage in the region 6.5h < R.A. < 9.5h, $25^{\circ}$ < Dec. < $39^{\circ}$ \citep{2017AandA...598A..78I}. This survey, observed at 150 MHz frequency with a resolution of $25\times25\text{arcsec}^2$ (or $25^{"}\times 25^{"}/\text{cos}(\text{Dec} - 19^{\circ}$) for sources at Dec < $19^{\circ}$, complements the broad frequency range and surface-brightness sensitivity of the MWA. The survey has a sensitivity limit below 5 mJy/beam ($7\sigma$ threshold) over the majority of its coverage and an astrometric accuracy of < 2 arcsec in R.A. and Dec.

\subsubsection{SUMSS catalogue and images (843 MHz)}
The Molonglo Observatory Synthesis Telescope (MOST; \citealt{1981PASAu...4..156M,1991AJp}) previously surveyed the southern sky (Dec. < $-30^{\circ}$, |b| > $10^{\circ}$) at 843 MHz creating SUMSS with a resolution of $45^{"}\times 45^{"}$. At Dec. < $-50^{\circ}$ SUMSS reaches a sensitivity limit of $\sim 5\sigma$ (6 mJy/beam) threshold and 10 mJy/beam at Dec. > $-50^{\circ}$. For sources brighter than 200 mJy at 843 MHz, the observed astrometric accuracy is $\sim$ 1-2 arcsec, while the largest positional error observed for the survey is $\sim$ 30 arcsec. This survey has a similar resolution and sensitivity to NVSS. For this work, SUMSS is used for GLEAM components at Dec $< -39.5^{\circ}$ and NVSS for GLEAM components at Dec $\geq -39.5^{\circ}$.

\subsubsection{NVSS catalogue and images (1.4 GHz)}
NVSS constitutes observations from the Very Large Array (VLA; \citealt{1980ApJS...44..151T}) which surveyed the northern sky (Dec. = $-40^{\circ}$) at 1.4 GHz frequency. The survey has a resolution of 45 arcsec and a $5\sigma$ threshold in peak source brightness of $\sim$2.5 mJy/beam. For sources brighter than 15 mJy, the observed astrometric accuracy is $\leq$ 1 arcsec.

\subsection{Mid-infrared}
\subsubsection{AllWISE catalogue}
WISE, launched in 2009, is an infrared space telescope that has mapped the entire sky at 4 wavelength bands 3.4, 4.6, 12 and 22 $\mu m$ known as W1, W2, W3 and W4, respectively. The 4 bands correspond to 6.1, 6.4, 6.5, and 12.0 arcsec resolution. In this work, we use the 6.1 arcsec resolution images (W1 band, 3.4 $\mu m$) as our grayscale and the AllWISE \citep{2013wise.rept....1C} positions for host galaxy identification, complemented by the radio contours from the surveys mentioned in section \ref{subsec:radio_data}. At the $5\sigma$ threshold, AllWISE has an improved sensitivity of 0.054, 0.071, 0.73, and 5.0 mJy in the four WISE bands, respectively, and positional error of < 1 arcsec, as compared to the WISE All-Sky data release \citep{2012wise.rept....1C}. 

\subsection{Optical redshift surveys}
\subsubsection{6dFGS}
The 6-degree Field Galaxy Survey (6dFGS; \citealt{2004MNRAS.355..747J}) obtained optical spectroscopy for the southern sky (Dec < $0^{\circ}$, |b| > $0^{\circ}$) using the 6dF multifibre spectroscopy on the United Kingdom Schmidt Telescope (UKST; \citealt{tritton_1978}). For sources brighter than $K$ = 12.65 in the 2MASS Extended Source Catalog \citep{2000AJ....119.2498J}, the final data release of 6dFGS (DR3; \citealt{2009MNRAS.399..683J}) provides redshifts, which we use for our work. 6dFGS has a resulting median redshift of 0.053.

\subsubsection{2MRS}
The 2 Micron All Sky Survey (2MASS; \citealt{2006AJ....131.1163S}) Redshift Survey (2MRS; \citealt{2012ApJS..199...26H}) constitute spectroscopic observations from various telescopes (see Table \ref{tab:table1}). The survey initially observed galaxies with $K_s$ < 11.25 mag and later observed galaxies below the Galactic latitude limit of 6dFGS (|b| > $10^{\circ}$). We use redshifts from this survey for G4Jy sources not detected in the 6dFGS. For sources which had no available redshift in 6dFGS and 2MRS, we further searched their redshifts from the NASA/IPAC Extragalactic Database, Kilo-Degree Survey (KiDS; \citealt{2017AandA...604A.134D}, Sloan Digital Sky Survey (SDSS; \citealt{2015ApJS..219...12A}, and Half Million Quasars (HMQ) Catalogue \citep{2015PASA...32...10F}.
\begin{table}
    \centering
    \caption{Telescopes and instruments used to compile 2MRS. "Res." in column five is the spectral resolution. Table taken from \citet{2012ApJS..199...26H}. }
    \begin{tabular}{p{0.3\linewidth}p{0.1\linewidth}p{0.15\linewidth}p{0.05\linewidth}p{0.22\linewidth}} \hline
Observatory/telescope & Camera	& Coverage & Res. & No. of galaxies with $K_s$ \\ \hline
  &  & (\AA) & (\AA) & <11.75 $\,$ >11.75 \\ \hline
Fred L. Whipple 1.5 m & FAST & 3500-7400 & 5 & 7590$ \;\;\;\;$ 2596 \\
Cerro Tololo	1.5 m & RCSpec & 3700–7200 & 7 & 3245 $\;\;\;\;$ 238 \\
McDonald	2.1 m & es2 & 3700–6400 & 4 & 114 $\;\;\;\;\;\;$  50 \\
Cerro Tololo	4 m & RCSpec & 3700–7400 & 3 & 48	 \\
Hobby–Eberly	9.2 m & LRS & 4300–10800 & 9 & 3  \\ \hline
    \end{tabular}
    \label{tab:table1}
\end{table}

\section{Method}\label{sec:method}
\subsection{Construction of the MeerKAT-2019 subset}
Our sample, defined as the MeerKAT-2019 subset, comprises 140
radio sources from the G4Jy Sample. Poor resolution data at 25 and 45 arcsec and complex morphologies limited host-galaxy identification. The MeerKAT-2019 subset includes; 13 sources (Table \ref{tab:13sources}) whose host galaxy was provided \citep{2020PASA...37...17W, 2020PASA...37...18W}, one G4Jy source (G4Jy 1523) whose host galaxy is affected by nearby bright, mid-infrared emission, and 126 sources with uncertain identification. For the 13 sources, there were some discrepancies in the literature concerning their host galaxy; therefore, higher-resolution images were required to confirm the provided host galaxy. \\
\begin{table}
    \centering
    \caption{A list of sources in the MeerKAT-2019 subset that needed higher resolution images to confirm the host-galaxy identification provided by \citet{2020PASA...37...17W, 2020PASA...37...18W}.}
    \begin{tabular}{c|c} \hline
    G4Jy name & GLEAM component name(s) \\ \hline
        G4Jy 40 &  GLEAM J002056-190853 and GLEAM J002112-191041 \\
        G4Jy 285 & GLEAM J024103+084523 and GLEAM J024107+084452\\
        G4Jy 333 & GLEAM J031152-312959  \\
        G4Jy 570 & GLEAM J054049-614233  \\
        G4Jy 717 & GLEAM J083710-195152 \\
        G4Jy 747 & GLEAM J090147-255516 \\
        G4Jy 917 & GLEAM J112554-352321 \\
        G4Jy 1094 & GLEAM J134855-252700 \\
        G4Jy 1205 & GLEAM J145509-365543  \\
        G4Jy 1260 & GLEAM J152659-135059  \\
        G4Jy 1537 & GLEAM J192606-573954 \\
        G4Jy 1638 & GLEAM J203444-354849 \\
        G4Jy 1741 & GLEAM J215415-455319 and GLEAM J215435-454954 \\
        \hline
    \end{tabular}
    \label{tab:13sources}
\end{table}

\subsection{Overlays}
We use APLpy \citep{2012ascl.soft08017R} to create overlays that are $\sim$10 arcmin across for all the sources in the MeerKAT-2019 subset. The overlays consist of contours from four surveys (GLEAM, NVSS/SUMSS, TGSS and MeerKAT) overlaid onto the mid-infrared WISE (W1, $3.4~\mu m$) images. The W1 band is used as our grayscale image because it has better sensitivity and resolution than W2, W3, and W4 bands. However, for radio sources with no detected host in mid-infrared, we use the optical Pan-STARRS (Panoramic Survey Telescope and Rapid Response System; \citealt{chambers2019panstarrs1}) images instead. Radio sources with a faint infrared host are most likely at high redshifts. So we need deep surveys to probe them, hence the use of Pan-STARRS images in this study. We set the lowest contour level to $3\sigma$ for the radio surveys in the overlays. However, for sources with artefacts in the MeerKAT imaging, we set the lowest contour to a higher value. We then cross-matched the MeerKAT-2019 subset with the AllWISE catalogue, using a radius of 3 arcmin from the centroid position via TOPCAT \citep{2005ASPC..347...29T}, and overplot these AllWISE positions in the overlays (indicated by green crosses, 'x'). Figure \ref{fig:g4jy40} is an example of a 10 by 10 arcmin overlay showing all of the datasets mentioned in this section.

\subsection{Radio morphology classification}
Host galaxy identification is highly associated with understanding the radio morphology of the radio source. In this paper, we determine the radio morphology of the MeerKAT-2019 subset using the $\sim7$ arcsec resolution images from MeerKAT, and use the following four categories:
\begin{itemize}
    \item `single' - the radio source has a compact morphology in the MeerKAT image,
    \item `double' - the radio source has two distinct lobes in the MeerKAT image but there is no detection of the radio core, or the source has an extended, elongated emission which is suggestive of radio lobes in the MeerKAT image,
    \item `triple' - the radio source has a radio core and two distinct radio lobes in the MeerKAT image,
    \item `complex' - the radio source has a morphology that does not meet the above three categories. 
\end{itemize}
\vspace{0.3cm}
\subsection{Host galaxy identification and flags}
We aim to identify the host galaxy of the radio emission by assessing the radio morphology and determining which AllWISE position is likely the host galaxy. For sources with `single' morphology, the host galaxy is likely to be at the centre. The host galaxy is likely to be located between the two distinct radio lobes for sources with `double' morphology. However, if there is more than one AllWISE position on-axis, we leave the source unidentified; unless there is a definitive core position. For radio sources with `triple' morphology, we expect the host galaxy to coincide with the position of the radio core (see Figure \ref{fig:g4jy40}). Lastly, we are unable to provide the host galaxy for sources with `complex' morphology. Following \cite{2020PASA...37...17W,2020PASA...37...18W}, we then use the following labels to indicate the host flag for each source in the MeerKAT-2019 subset;
\begin{itemize}
    \item `i' - the source has a corresponding host galaxy identified in the AllWISE/2MASS catalogue,
    \item `u' - it is unclear which AllWISE or 2MASS source is the host galaxy as (1) the source has complex radio morphology or (2) there is no radio core detected in the MeerKAT image that coincides with the AllWISE/2MASS position,
    \item `m' - either the source is affected by nearby bright mid-infrared emission, or there is no corresponding AllWISE position,
    \item `n' - given the type of radio emission involved (e.g., cluster relic or radio halo), no host-galaxy position should be indicated.
\end{itemize}

\subsection{Flux densities at 1.3 GHz}
We calculated the integrated flux density at 1.3 GHz ($S_{\rm{int}, 1.3~\rm{GHz}}$) for 138\footnote{\label{note0}Of the 140 sources, we were unable to obtain images for two sources (G4Jy 453 and G4Jy 456) as the visibilities were entirely flagged due to a burst of strong RFI.} G4Jy sources by summing the flux densities within the $3\sigma$ contour level, where $\sigma$ is the local rms in the 1.3 GHz image from MeerKAT. This integrated flux density is in Jy/beam, and to convert it to Jy; we divide the summation by the beam area. We provide the core intensity ($S_{\rm{core}}, 1.3~\rm{GHz}$), in Jy/beam units, for sources with an identified host galaxy by extracting the flux density at the pixel value of the host galaxy position in the MeerKAT image. Both integrated flux densities and core intensity are provided in Table \ref{tab:table5}.

\subsection{Calculation of spectral indices}
We calculated the two-point spectral index ($\alpha^{1300~\rm{MHz}}_{151~\rm{MHz}}$) between 151 and 1300 MHz for the MeerKAT-2019 subset using equation \ref{eq_alpha};

\begin{equation}
    \alpha^{1300~\rm{MHz}}_{151~\rm{MHz}} = \frac{\rm{log}_{10}(S_{\nu_2}) - \rm{log}_{10}(S_{\nu_1})}{\rm{log}_{10}(\nu_2) - \rm{log}_{10}(\nu_1)} \label{eq_alpha}
\end{equation}

where $S_{\nu_1}$ and $S_{\nu_2}$ are the integrated flux densities at $\nu_1 = 151~\rm{MHz}$ and $\nu_2 = 1300~\rm{MHz}$, respectively. The integrated flux densities at 151 MHz were obtained from the G4Jy catalogue \citep{2020PASA...37...17W, 2020PASA...37...18W}. The spectral index values are presented in Table \ref{tab:table5}.

\section{Results}\label{sec:results}
The host-galaxy identification of radio emission of 140 G4Jy sources relies upon the $\sim$7-arcsec resolution images from MeerKAT as existing radio images from TGSS and NVSS/SUMSS with resolutions of 25 and 45 arcsec, respectively, do not provide sufficient detail. We resolve the 'complex' morphologies evident in the TGSS and/or NVSS/SUMSS images with these new images from MeerKAT and enable host-galaxy identification for 98 sources in the MeerKAT-2019 subset, through visual inspection of the overlays and manually pinpointing the host galaxy. Of the 98 sources with an identified host galaxy in the MeerKAT-2019 subset, the host galaxies of two sources (G4Jy 1371 and G4Jy 1540) were identified using the optical Pan-STARRS images as the host galaxies of these sources are not detected in the mid-infrared AllWISE images. In Table \ref{tab:table5}, we present our findings for the MeerKAT-2019 subset. 

\begin{figure*}
    \centering
\includegraphics[width=0.5\textwidth]{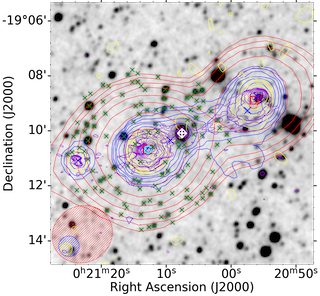}
\caption{An overlay of G4Jy 40, centred at R.A. = 00:21:07.53, Dec. = -19:10:05.4. The radio contours are from GLEAM (170 - 231 MHz; red), NVSS (1.4 GHz; blue), TGSS (150 MHz; yellow) and MeerKAT (1.3 GHz; purple), overlaid on the inverted grayscale WISE (3.4 $\mu m$) image. Shown in the bottom left corner is the beam size for each survey; GLEAM (red), NVSS(blue), TGSS (yellow) and MeerKAT (purple). The cyan hexagon indicates the brightness-weighted centroid position, the green crosses `x' signs are the AllWISE positions within 3 arcmin from the centroid position, and the white `+' sign is the AllWISE position for the corresponding host galaxy for this source. The white diamond is the 6dFGS position, the red squares are the GLEAM components positions, and blue crosses are NVSS/SUMSS components positions.}
\label{fig:g4jy40}
\end{figure*}

\begin{table*}
\caption{Properties of sources in the MeerKAT-2019 subset. Columns 1-3 are; the G4Jy name, host flag, and the host galaxy name. Columns 4 is the redshift, column 5 is the redshift reference, and column 6 is the spectral index ($\alpha$) between 151 and 1300 MHz. Columns 7-9 are the core intensity at 1300 MHz (SC13) in mJy/beam, integrated flux density at 1300 MHz (S13) in Jy, and integrated flux density at 1400 MHz (S14) in Jy. Columns 10-11 are the radio luminosities at 1300 MHz (L13) and 1400 MHz (L14) in W/Hz. Column 12 is the linear size (LS), and column 13 is the radio morphology label based on MeerKAT $\sim7-\text{arcsec}$ resolution images. The full table is available online.} 
\label{tab:table5}
 \begin{tabular}{nLcLLLccLccLc}
  \hline
G4Jy name & Host flag & Host name & z & z ref. & $\alpha$ & $SC13$ &  $S13$ & $S14$ & $L13$ & $L14$ & LS & Morphology \\ 
& & & & & & & & & & & & \\
& & & & & & [mJy/beam] & [Jy] & [Jy] & [W/Hz] & [W/Hz] & [Mpc] & \\
  \hline
G4Jy 14 & i & WISEA J000707.12$+$053609.6 & 0.216 & (1) &  & $98.166$ & $0.441{\pm}0.085$ &  &  &  &  & Double \\
G4Jy 40 & i & WISEA J002107.53$-$191005.4 & 0.096 & (2) & $-0.707$ & $0.230$ & $1.261{\pm}0.140$ &  & $2.85{\times} 10^{25}$ &  &  & Triple \\
G4Jy 47 & i & WISEA J002531.45$-$330246.2 & 0.05 & (2) & $-0.815$ & $0.424$ & $1.527{\pm}0.162$ & 1.494 & $8.95{\times}10^{24}$ & $8.75{\times} 10^{24}$ & 0.091  & Double\\
G4Jy 77 & u &  &  &  & $-2.604$ &  & $0.046 \pm 0.027$ & 0.035 &  &  &  & Complex \\
G4Jy 84 & i & WISEA J004605.01$-$633319.2 & 0.075 & (2) & $-0.82$ & $0.269$ & $0.731{\pm}0.100$ &  & $9.94{\times} 10^{24}$ &  & 0.215  & Triple \\
		\hline
	\end{tabular}
\end{table*}

\subsection{Radio sources with an identified host galaxy}
\subsubsection{Previous identifications}\label{sec:13sources}
\begin{figure*}%
\begin{multicols}{2}
\begin{subfigure}{\columnwidth}
\centering
\includegraphics[width=0.8\columnwidth]{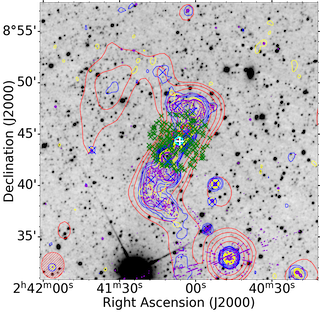}\par\caption{G4Jy 285}\label{fig:g4jy_285}
\end{subfigure}%

\begin{subfigure}{\columnwidth}
\centering
\includegraphics[width=0.85\columnwidth]{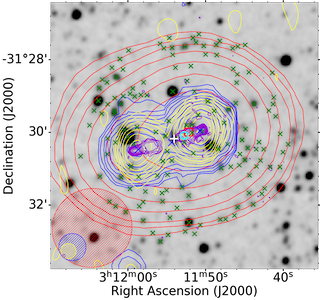}\par\caption{G4Jy 333}\label{fig:g4jy_333}
\end{subfigure}%
\end{multicols}

\begin{multicols}{2}
\begin{subfigure}{\columnwidth}
\centering
\includegraphics[width=0.85\columnwidth]{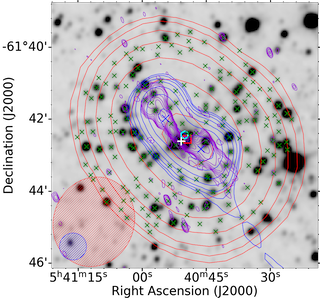}\par\caption{G4Jy 570}\label{fig:g4jy_570}
\end{subfigure}%

\begin{subfigure}{\columnwidth}
\centering
\includegraphics[width=0.85\columnwidth]{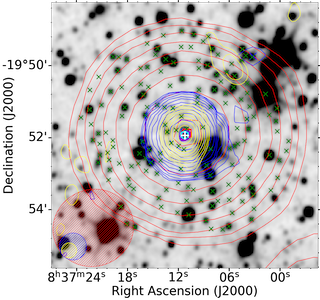}\par\caption{G4Jy 717}\label{fig:g4jy_717}
\end{subfigure}
\end{multicols}

\begin{subfigure}{\columnwidth}
\centering
\includegraphics[width=0.85\columnwidth]{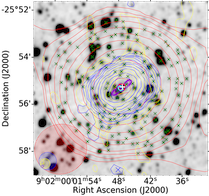}\par\caption{G4Jy 747}\label{fig:g4jy_747}%
\end{subfigure}

\caption{Overlays for 5 of 13 sources in the G4Jy Sample for which the literature has disagreements regarding the host galaxy (see section \ref{sec:13sources}). The datasets, symbols, beams and contours are the same as those described in Figure \ref{fig:g4jy40}. }
\label{fig:identified1}
\end{figure*}

\cite{2020PASA...37...17W, 2020PASA...37...18W} provided the host-galaxy identification for 13 of 140 G4Jy radio sources (Figures \ref{fig:g4jy40}, \ref{fig:identified1} and \ref{fig:identified2}). However, as these were `borderline' identifications (with some accompanied by discrepancies in the literature), higher-resolution data was needed to confirm the host galaxy. With the MeerKAT follow-up images, we confirm the host galaxy of the 13 sources provided by \citet{2020PASA...37...17W, 2020PASA...37...18W}.\\

\noindent
\textbf{G4Jy 40} (Figure \ref{fig:g4jy40}): \cite{2020PASA...37...17W, 2020PASA...37...18W} inferred the host galaxy (AllWISE J002107.53$-$191005.4) of this radio source (with radio lobes GLEAM J002056$-$190853 and GLEAM J002112$-$191041) based on the interpretation of \citet{1998AandAS..132...31N} that G4Jy 40 has a `double' morphology with the largest angular size of 252 arcsec, using the radio map of \citet{1975MmRAS..79....1S}. However, the 4.7 GHz image from \citet{Reid_1999} reveals a radio source with `triple' morphology that spans $\sim$160 arcsec from lobe to lobe (cyan contours in Figure 15 of \citealt{2020PASA...37...17W}). Due to the incorrect coordinates presented in the \cite{Reid_1999} image, a higher resolution image needs to be used to confirm if GLEAM J002056$-$190853 is associated with G4Jy 40. The $\sim$7-arcsec resolution image from MeerKAT  reveals a radio source with `triple' morphology. The tight concentric contours at the centre (R.A. = 00:21:05.54, Dec. = $-$19:10:05.34) verify that AllWISE J002107.53$-$191005.4 (g002107.53$-$191005.4, $z = 0.096$) is the mid-infrared host galaxy of the extended emission, and GLEAM J002056$-$190853 and GLEAM J002112$-$191041 are the radio lobes, with flux densities of $0.919323~\mathrm{Jy}$ and $4.862407~\mathrm{Jy}$ at 151 MHz, respectively. \\

\noindent
\textbf{G4Jy 285} (GLEAM J024103$+$084523 and GLEAM J024107$+$084452; Figure \ref{fig:g4jy_285}): The host galaxy of the extended radio emission of G4Jy 285, also known as NGC 1044 (4C $+08.11$), is AllWISE J024106.17$+$084416.9, also detected in 2MRS (J02410618$+$0844167, $z = 0.021$). This identification is in agreement with \citet{2020PASA...37...17W, 2020PASA...37...18W} and \citet{2012AandA...544A..18V}. The radio source traces a zig-zag emission; however, it was unclear whether the low-frequency emission towards the northeast (GLEAM J024133$+$084940) is associated with G4Jy 284 \citep{2020PASA...37...17W, 2020PASA...37...18W}. We cannot confirm whether the radio emission is associated with G4Jy 284 as there is no detection of this emission in our 1.3 GHz image from MeerKAT. We only have a shallow `snapshot' image. A longer observation time might reveal something.\\

\noindent
\textbf{G4Jy 333} (GLEAM J031152$-$312959; Figure \ref{fig:g4jy_333}): MeerKAT contours reveal `triple' morphology with edge-brightened lobes. We confirm that AllWISE J031154.01$-$313010.6 is the host galaxy. However, we find no redshift information in the literature. \\

\noindent
\textbf{G4Jy 570} (GLEAM J054049$-$614233; Figure \ref{fig:g4jy_570}): `Double' extended morphology is evident in SUMSS and MeerKAT contours. This radio source (PKS B0540$-$617) is in cluster A3362. The host galaxy of this radio source is AllWISE J054050.82$-$614237.2 ($z = 0.081$) which agrees with \cite{2020PASA...37...17W, 2020PASA...37...18W} and the optical identification of \citet{1992ApJS...80..137J}.  \\

\noindent
\textbf{G4Jy 717} (GLEAM J083710$-$195152; Figure \ref{fig:g4jy_717}): This radio source, known as PKS B0834$-$19, has `single' morphology evident in all surveys. We confirm that the host galaxy of this source is AllWISE J083711.18$-$195156.6 ($z = 1.032$), which agrees with the optical identification provided by \cite{1988AandAS...75..173F} and \cite{1994MNRAS.269..998D}. We note the artefacts in the MeerKAT image, which result in an overestimated integrated flux density at 1.3 GHz. We, therefore, present an overlay where the lowest contour level has been adjusted to $48\sigma$. G4Jy 717 has the flattest spectral index ($\alpha^{1300~\rm{MHz}}_{151~\rm{MHz}} = -0.17$) in the MeerKAT-2019 subset. \\

\noindent
\textbf{G4Jy 747} (GLEAM J090147$-$255516; PKS 0859$-$25, Figure \ref{fig:g4jy_747}): `Double' extended radio morphology is evident in the MeerKAT contours. We confirm that AllWISE J090147.26$-$255516.2 ($z = 0.305$) is the host galaxy of the radio emission. The MeerKAT contours start at $48\sigma$ as artefacts are present at lower levels.\\

\begin{figure}
\begin{subfigure}{\columnwidth}
\centering
\includegraphics[width=0.85\columnwidth]{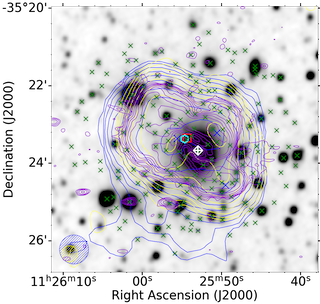}\par\caption{An overlay of G4Jy 917}\label{fig:g4jy_917}
\end{subfigure}

\begin{subfigure}{\columnwidth}
\centering
\includegraphics[width=0.9\columnwidth]{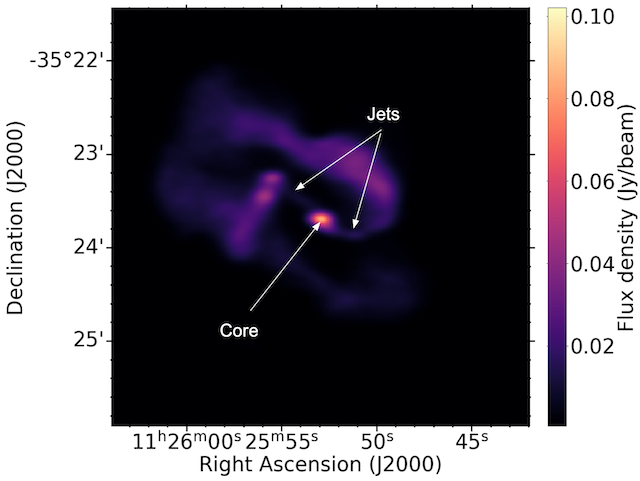}\par\caption{MeerKAT intensity map.}\label{fig:g4jy_917intensity_map}
\end{subfigure}

\caption{Top panel: An overlay of G4Jy 917 (one of the 13 sources with discrepancies in the literature concerning the host galaxy identification). The datasets, symbols, beams and contours are the same as those described in Figure \ref{fig:g4jy40}. Bottom panel: The intensity map at 1.3 GHz of G4Jy 917 showing the structure of this radio source. The morphology is very complex, and includes the emission eastwards of the core forming a right-angled bend.}
\end{figure}

\begin{figure*}
\begin{multicols}{2}
\begin{subfigure}{\columnwidth}
\centering
\includegraphics[width=0.83\columnwidth]{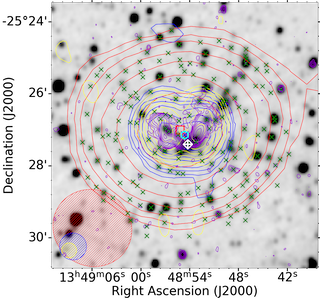}\par\caption{G4Jy 1094}\label{fig:g4jy_1094}
\end{subfigure}%

\begin{subfigure}{\columnwidth}
\centering
\includegraphics[width=0.83\columnwidth]{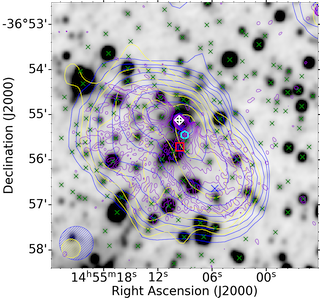}\par\caption{G4Jy 1205}\label{fig:g4jy_1205}
\end{subfigure}
\end{multicols}

\begin{multicols}{2}
\begin{subfigure}{\columnwidth}
\centering
\includegraphics[width=0.83\columnwidth]{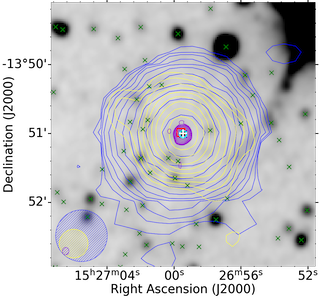}\par\caption{G4Jy 1260}\label{fig:g4jy_1260}
\end{subfigure}

\begin{subfigure}{\columnwidth}
\centering
\includegraphics[width=0.83\columnwidth]{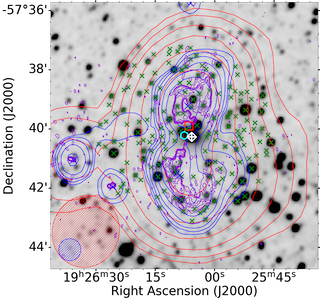}\par\caption{G4Jy 1537}\label{fig:g4jy_1537}
\end{subfigure}%
\end{multicols}

\begin{multicols}{2}
\begin{subfigure}{\columnwidth}
\centering
\includegraphics[width=0.83\columnwidth]{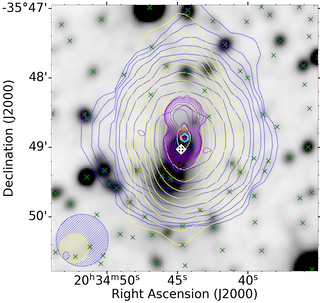}\par\caption{G4Jy 1638}\label{fig:g4jy_1638}
\end{subfigure}%

\begin{subfigure}{\columnwidth}
\centering
\includegraphics[width=0.84\columnwidth]{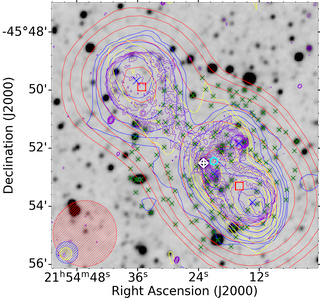}\par\caption{G4Jy 1741}\label{fig:g4jy_1741}
\end{subfigure}
\end{multicols}

\caption{Overlays for 6 of 13 sources in the G4Jy Sample for which the literature has disagreements regarding the host galaxy (see section \ref{sec:13sources}). The datasets, symbols, beams and contours are the same as those described in Figure \ref{fig:g4jy40}. }
\label{fig:identified2}
\end{figure*}

\noindent
\textbf{G4Jy 917} (GLEAM J112554$-$352321): PKS B1123$-$315 (Figure \ref{fig:g4jy_917}) is in the cluster AS 665. The higher resolution image from MeerKAT shows the distinct inner structure of this radio source where knots are evident in the eastwards jet (Figure \ref{fig:g4jy_917intensity_map}). Furthermore, this jet is bent forming a right angle while the emission resulting from the westwards jet appears to be pushed towards south. The radio core is evident in the 1.3 GHz image from MeerKAT, which coincides with the position of the radio core from the 4.9 GHz image provided by \citet{1989MNRAS.236..737E}. The host galaxy of the radio emission is AllWISE J112552.95$-$352340.3, which appears in 6dFGS (g1125529$-$352340, z = 0.033). This identification is in agreement with \cite{2012AandA...544A..18V}, and a deeper MeerKAT image will be presented by Thorat et al. (in prep.).\\

\noindent
\textbf{G4Jy 1094}: The wide-angle tail (WAT) radio morphology is evident in the MeerKAT image (Figure \ref{fig:g4jy_1094}). We confirm that the radio source is hosted by the mid-infrared AllWISE source J134854.17$-$252724.5, detected in 6dFGS at $z = 0.126$. This radio source is in the cluster Abell 1791, located at the centre \citep{1989AuJPh..42..633S}. \\

\noindent
\textbf{G4Jy 1205}: The radio source known as PKS 1452$-$367 (Figure \ref{fig:g4jy_1205}) has a `triple' morphology with diffuse edges evident in MeerKAT contours. We confirm that the mid-infrared AllWISE J145509.61$-$365507.4 (g1455096$-$365508, $z = 0.095$) is the host galaxy, which is in agreement with \citet{2012AandA...544A..18V}. \\

\noindent
\textbf{G4Jy 1260}: The radio source known as PKS B1526$-$136 (Figure \ref{fig:g4jy_1260}) has a `single' morphology evident in all surveys. We confirm that the host galaxy is the mid-infrared AllWISE J152659.45$-$135100.0 at redshift $z = 1.687$. This radio-loud quasar has the highest redshift in the MeerKAT-2019 subset (so far) and is one of the sources with a very flat spectral index ($\alpha_{1300~\rm{MHz}}^{151~\rm{MHz}} = -0.36$). The MeerKAT contours start at $24\sigma$ as artefacts are present at lower levels. \\

\noindent
\textbf{G4Jy 1537} (PKS B1921$-$577 Figure \ref{fig:g4jy_1537}): `Triple' morphology is evident in the 1.3 GHz image from MeerKAT. We confirm that the mid-infrared AllWISE J192605.75$-$574016.4 (6dFGS g1926057$-$574017, $z = 0.061$) is the host galaxy of the WAT radio source. The MeerKAT contours start at $12\sigma$ as artefacts are present at lower levels. \\

\noindent
\textbf{G4Jy 1638} (Figure \ref{fig:g4jy_1638}): The head-tail radio morphology is evident in the 1.3 GHz image from MeerKAT. The corresponding host galaxy of the radio emission is AllWISE J203444.74$-$354901.7, appearing in 6dFGS as g2034447$-$354902 ($z = 0.089$). \\

\noindent
\textbf{G4Jy 1741} (Figure \ref{fig:g4jy_1741}): Edge-brightened radio lobes and a compact core are evident in the 1.3 GHz image from MeerKAT. We confirm that AllWISE J215422.96$-$455231.3 is the host galaxy of the `triple' morphology radio source, detected in 6dFGS (g2154230$-$455232, $z = 0.145$).

\subsubsection{Newly identified X- and S-/Z-shaped radio sources}
\begin{figure*}
\begin{multicols}{2}
\begin{subfigure}{\columnwidth}
\centering
\includegraphics[width=0.83\columnwidth]{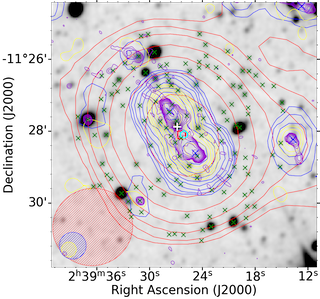}
\par\caption{G4Jy 284}\label{fig:g4jy284} 
\end{subfigure}

\begin{subfigure}{\columnwidth}
\centering
\includegraphics[width=0.83\columnwidth]{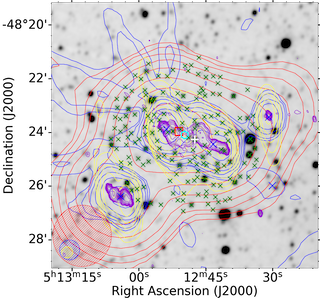}\par\caption{G4Jy 530}\label{fig:g4jy530}  
\end{subfigure}
\end{multicols}

\begin{multicols}{2}
\begin{subfigure}{\columnwidth}
\centering
\includegraphics[width=0.83\columnwidth]{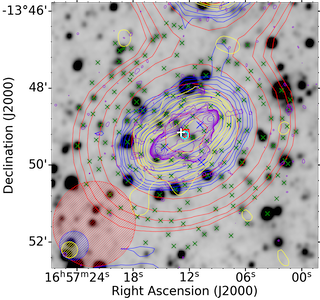}\par\caption{G4Jy 1377}\label{fig:g4jy1377}  
\end{subfigure}

\begin{subfigure}{\columnwidth}
\centering
\includegraphics[width=0.83\columnwidth]{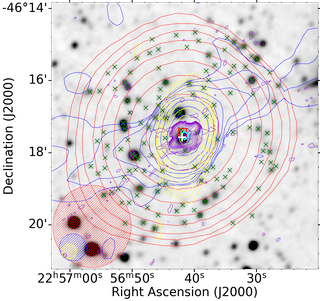}\par\caption{G4Jy 1798} \label{fig:g4jy1798} 
\end{subfigure}
\end{multicols}

\begin{subfigure}{\columnwidth}
\centering
\includegraphics[width=0.83\columnwidth]{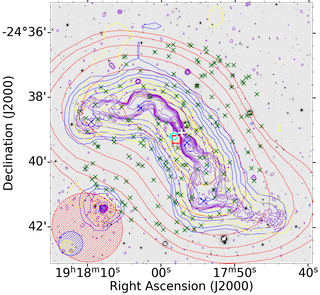}\par\caption{G4Jy 1523}\label{fig:g4jy1523}
\end{subfigure}

\caption{Four radio sources with X-shaped morphology and one G4Jy source (G4Jy 1523) with S-/Z-shaped morphology in the MeerKAT-2019 subset. For G4Jy 1523, PanSTARRS image is used as grayscale base for the overlay and green crosses are from 2MASS. The datasets, symbols, beams and contours are the same as those described in Figure \ref{fig:g4jy40}.}
\label{fig:XSZ}
\end{figure*}

X- and S-/Z-shaped radio galaxies form a subclass of radio galaxies with intriguing radio morphology. These radio galaxies are characterised by an additional pair of secondary radio lobes (generally referred to as wings) that are more diffuse, have low surface brightness and are misaligned from the radio core compared to the primary lobes \citep{1978Natur.276..588E, 1992ersf.meet..307L}. The orientation of the secondary lobes generally defines the X- or S-/Z-shaped morphology. For S-/Z-shaped morphology, the low surface brightness secondary lobes emerge from the edges of the high surface brightness secondary lobes. For X-shaped morphology, the axis of the secondary lobes is nearly perpendicular to the axis of the primary lobes. In the MeerKAT-2019 subset, four radio sources (G4Jy 284, G4Jy 530, G4Jy 1377 and G4Jy 1798) have an X-shaped morphology, and one radio source (G4Jy 1523) has an S-/Z-shaped morphology (Figure \ref{fig:XSZ}). Below, we discuss only a few (11) that are interesting. \\

\noindent
\textbf{G4Jy 284} (GLEAM J023926$-$112806, Figure \ref{fig:g4jy284}): A (borderline) X-shaped morphology, and the detection of the radio core, is evident in the MeerKAT contours. The mid-infrared source that coincides with the radio core is AllWISE J023926.84$-$112752.6. However, there is no redshift information in the literature for the host galaxy. \\

\noindent
\textbf{G4Jy 530} (GLEAM J051250$-$482358, PKS 0511-48, Figure \ref{fig:g4jy530}): For this radio source, \cite{1985MNRAS.212..809S}, using a radio map with a spatial resolution of $53"\times 46"$, reported an optical identification that matches AllWISE J051247.41$-$482416.5, classified as a Seyfert type 2 galaxy. However, \cite{2020PASA...37...17W, 2020PASA...37...18W} asserted that the radio map of \cite{1985MNRAS.212..809S} is not of sufficient resolution to rule out the presence of another mid-infrared source on the axis connecting the radio lobes. The X-shape morphology evident in the MeerKAT contours support the optical identification of \cite{1985MNRAS.212..809S}, and we, therefore, identify AllWISE J051247.41$-$482416.5 as the host galaxy. The low surface brightness radio emission (secondary lobes) might be due to plasma backflow in the primary lobes. The MeerKAT contours start at $12\sigma$ as artefacts are present at lower levels.\\

\noindent
\textbf{G4Jy 1377} (GLEAM J165712$-$134911, Figure \ref{fig:g4jy1377}): This radio source is known as PKS B1654-137. An X-shaped morphology is evident in the MeerKAT contours, with the primary lobes aligned northwest and southeast, while the secondary lobes are aligned northeast and southwest. The mid-infrared counterpart of this radio source is the AllWISE source, J165712.85$-$134909.5, at z = 0.124. \\

\noindent
\textbf{G4Jy 1798} (GLEAM J225641$-$461726, Figure \ref{fig:g4jy1798}): There are two AllWISE positions near the centroid, and it was not clear which AllWISE position is the most-likely host galaxy. The MeerKAT image reveals an X-shaped galaxy with the primary lobes aligned northwest and southeast while the secondary lobes are aligned northeast and southwest. We then use a DSS image as our grayscale and identify the mid-infrared host galaxy as 2MASS J22564196$-$4617345, at z = 0.080. \\

\textbf{G4Jy 1523} (GLEAM J191757$-$243917, Figure \ref{fig:g4jy1523}): MeerKAT contours reveal a Z-shaped radio galaxy suggestive of precessing jets. The radio source was previously assigned host flag `m' as the host galaxy in the AllWISE image is obscured by the nearby star \citep{2020PASA...37...17W, 2020PASA...37...18W}. As a result, the 2MASS image was used as the grayscale base image for the overlay, and the identified host galaxy is 2MASS 19175722$-$2439053. We find no redshift information in the literature, but follow-up SALT (Southern African Large Telescope; \citealt{2006SPIE.6267E..0ZB}) spectroscopy (PI: White) has been obtained for this source.

\subsubsection{Newly identified head-tail and wide-angle tail (WAT) radio sources}
Head-tail and WAT radio galaxies are a subclass of radio galaxies associated with clusters. The WAT and head-tail morphology observed is explained by the radio galaxy falling into a cluster. The ram pressure from the surrounding medium pushes the radio jets and lobes backwards, forming a WAT or head-tail morphology. In the MeerKAT-2019 subset, we find that 10 sources have head-tail morphology and 14 have WAT morphology.\\

\begin{figure*}
\begin{multicols}{2}
\begin{subfigure}{\columnwidth}
\centering
\includegraphics[width=0.83\columnwidth]{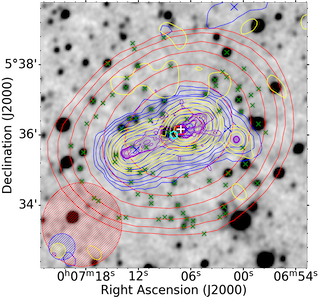}
\par\caption{G4Jy 14}\label{fig:g4jy14} 
\end{subfigure}

\begin{subfigure}{\columnwidth}
\centering
\includegraphics[width=0.83\columnwidth]{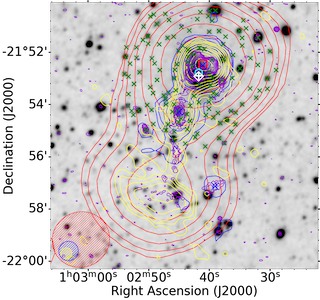}\par\caption{G4Jy 113}\label{fig:g4jy113}
\end{subfigure}
\end{multicols}

\begin{multicols}{2}
\begin{subfigure}{\columnwidth}
\centering
\includegraphics[width=0.83\columnwidth]{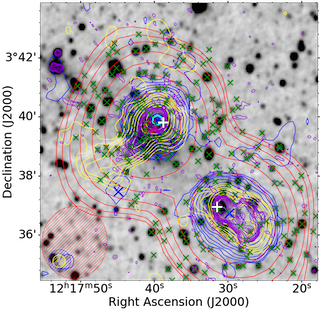}\par\caption{G4Jy 984 (towards NE) and G4Jy 983 (towards SW)}\label{fig:g4jy984} 
\end{subfigure}

\begin{subfigure}{\columnwidth}
\centering
\includegraphics[width=0.83\columnwidth]{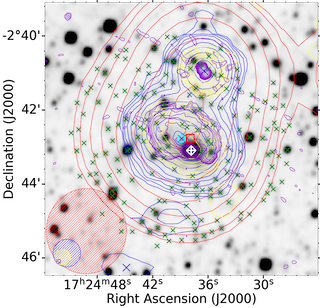}\par\caption{G4Jy 1410} \label{fig:g4jy1410} 
\end{subfigure}
\end{multicols}

\begin{subfigure}{\columnwidth}
\centering
\includegraphics[width=0.83\columnwidth]{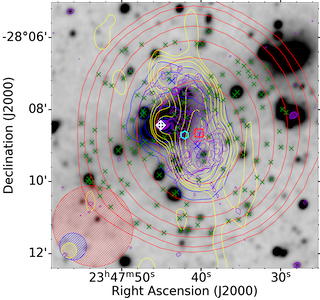}\par\caption{G4Jy 1852} \label{fig:g4jy1852} 
\end{subfigure}

\caption{Radio sources with head-tail morphology in the MeerKAT-2019 subset. The datasets, symbols, beams and contours are the same as those described in Figure \ref{fig:g4jy40}.}
\label{fig:head-tails}
\end{figure*}

\noindent
\textbf{G4Jy 14} (GLEAM J000707$+$053607, Figure \ref{fig:g4jy14}): `Double', extended radio morphology, suggestive of radio lobes, is evident in NVSS and TGSS contours. The host galaxy could not previously be provided for this source as there were two mid-infrared positions at a similar distance from the centroid position, and there was no detection of the radio core \citep{2020PASA...37...17W, 2020PASA...37...18W}. However, the radio contours from MeerKAT reveal two resolved sources, with one being a head-tail radio source towards the northeast. We, therefore, mark the two sources as unrelated (see subsection \ref{subsec:unrelated}) and identify the host galaxy of the head-tail radio source as the mid-infrared AllWISE J000707.12$+$053609.6 (z = 0.2161). The MeerKAT contours start at $12\sigma$ as artefacts are present at lower levels. \\

\noindent
\textbf{G4Jy 113} (GLEAM J010241$-$215227, Figure \ref{fig:g4jy113}): The radio source whose radio morphology from NVSS and TGSS is challenging to understand is located in the cluster Abell 133. \cite{2020PASA...37...17W, 2020PASA...37...18W} could not provide the host galaxy due to the disagreements in the literature concerning the radio morphology of G4Jy 113. \cite{Slee_2001} classified the radio source as a cluster relic generated via merger shocks, while \cite{Rizza_2000} and \cite{Fujita_2002} classified this source as a remnant radio lobe. Our radio image from MeerKAT indicates that G4Jy 113 is a head-tail radio source, with concentric radio-contours nearby highlighting an unrelated source. The former has spectral index, $\alpha^{1300~\rm{MHz}}_{151~\rm{MHz}} = -1.8$, and we identify the host galaxy as the AllWISE J010241.76$-$215254.2, detected in 6dFGS as g0102418$-$215256 (z = 0.057). This galaxy is referred to as galaxy 'H' by \cite{Slee_2001}. \\

\begin{figure*}
\begin{multicols}{2}
\begin{subfigure}{\columnwidth}
\centering
\includegraphics[width=0.8\columnwidth]{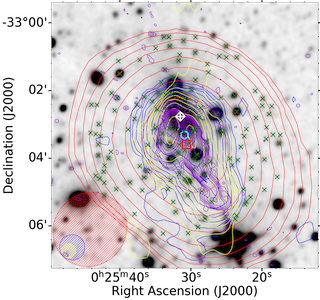}\par\caption{G4Jy 47} \label{fig:g4jy47} 
\end{subfigure}

\begin{subfigure}{\columnwidth}
\centering
\includegraphics[width=0.85\columnwidth]{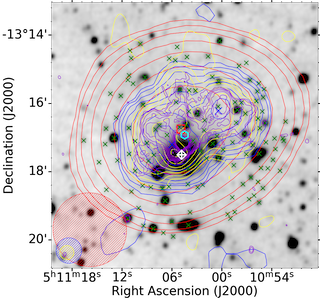}\par\caption{G4Jy 528} \label{fig:g4jy528} 
\end{subfigure}
\end{multicols}

\begin{multicols}{2}
\begin{subfigure}{\columnwidth}
\centering
\includegraphics[width=0.8\columnwidth]{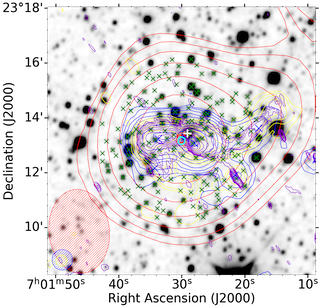}\par\caption{G4Jy 637} \label{fig:g4jy637} 
\end{subfigure}
    
\begin{subfigure}{\columnwidth}
\centering
\includegraphics[width=0.8\columnwidth]{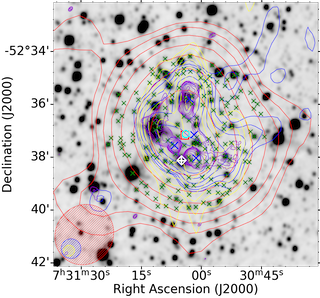}\par\caption{G4Jy 665} \label{fig:g4jy665} 
\end{subfigure}
\end{multicols}

\begin{multicols}{2}
\begin{subfigure}{\columnwidth}
\centering
\includegraphics[width=0.8\columnwidth]{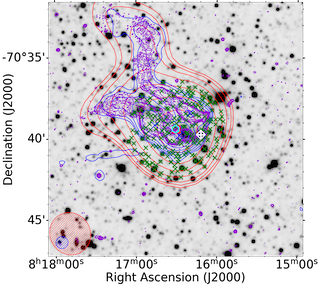}
\par\caption{G4Jy 693}\label{fig:g4jy693} 
\end{subfigure}

\begin{subfigure}{\columnwidth}
\centering
\includegraphics[width=0.85\columnwidth]{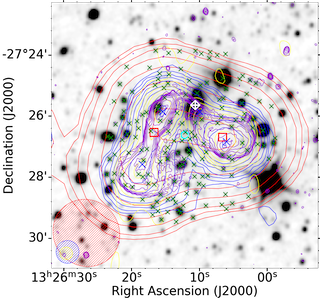}
\par\caption{G4Jy 1067} \label{fig:g4jy1067} 
\end{subfigure}
\end{multicols}

\caption{Radio sources with WAT radio morphology in the MeerKAT-2019 subset. The datasets, symbols, beams and contours are the same as those described in Figure \ref{fig:g4jy40}.}
\label{fig:wat}
\end{figure*}

\noindent
\textbf{G4Jy 984} (GLEAM J121740$+$033940; Figure \ref{fig:g4jy984}): There is a tail of emission evident in the TGSS contours for the northern source; however, it was unclear where the host galaxy could be. The MeerKAT contours reveal a far more detailed view of the structure of G4Jy 984, with the inner parts of the jet appearing distinct before combining into a tail towards the southeast. We identify the host galaxy as AllWISE J121738.77$+$033948.2 (z = 0.0778). The southwest radio source (not part of the MeerKAT-2019 subset) appears in the G4Jy catalogue as G4Jy 983 (4C +04.41), previously assigned host flag 'm'. This radio source is in the cluster Z5029 (R.A. = 12:17:14, Dec. = 03:39:23). Head-tail radio morphology is evident in the MeerKAT contours, and we identify the host galaxy as the mid-infrared AllWISE source J121731.43$+$033656.4 (z = 0.773).\\

\noindent
\textbf{G4Jy 1410} (GLEAM J172437$-$024246; Figure \ref{fig:g4jy1410}): Based on the compactness of the TGSS contours, it was unclear whether G4Jy 1410 (the southern source in the overlay) is a `double' or a head-tail galaxy, and as such, the host galaxy was not provided. MeerKAT contours reveal a head-tail radio galaxy (G4Jy 1410) and a `double' morphology radio galaxy (GLEAM J172436$-$024055) towards the north, which has a flux density of 3.79 Jy at 151 MHz \citep{2020PASA...37...17W, 2020PASA...37...18W}. Therefore, GLEAM J172436$-$024055 is not part of the G4Jy catalogue as it does not meet the criterion of flux density being above 4-Jy at 151 MHz. We identify the host galaxy of G4Jy 1410 as AllWISE J172437.79$-$024305.6 detected in 2MRS as J17243782$-$0243062 (z = 0.034).  This identification is in agreement with \citet{2012AandA...544A..18V}. \\

\noindent
\textbf{G4Jy 1852} (GLEAM J234740$-$280839; Figure \ref{fig:g4jy1852}): This radio source is in the cluster Abell 4035. A head-tail radio morphology with diffuse tail emission is evident in the MeerKAT contours. The mid-infrared source that coincides with the radio core is AllWISE J234745.06$-$280826.2. This identification corresponds to IC 5358, which is the brightest galaxy in a cluster. \\

\noindent
\textbf{G4Jy 47} (GLEAM J002530$-$330336; Figure \ref{fig:g4jy47}): Wide-angle tail radio morphology is evident in the higher-resolution image from MeerKAT. This radio source appears in the Molonglo Southern 4-Jy Sample (MS4; \citealt{Burgess_2006}) as MRC B0023-33. The host galaxy is the mid-infrared AllWISE J002531.45-330246.2 at redshift z = 0.050, which agrees with the host galaxy identification of \cite{Burgess_2006} and \citet{2012AandA...544A..18V}. An optical counterpart is ESO 350-G-15, the brightest galaxy in cluster AS 41 (ACO89). \\

\noindent
\textbf{G4Jy 528} (GLEAM J051104$-$131645; Figure \ref{fig:g4jy528}): WAT morphology with diffuse tail emission is evident in MeerKAT contours. The host galaxy is AllWISE J051104.82$-$131730.3 at $z = 0.043$.\\

\noindent
\textbf{G4Jy 637} (GLEAM J070130$+$231313; Figure \ref{fig:g4jy637}): A radio source known as 4C $+23.18$ in the literature \citep{1970AJ.....75..764O}. Based on the TGSS contours, \citet{2020PASA...37...17W, 2020PASA...37...18W} identified the morphology as WAT. However, it was unclear which AllWISE source was the host galaxy. With a higher resolution image from MeerKAT, we confirm that this is indeed a WAT radio source. We identify the host galaxy as AllWISE J070129.05$+$231325.6 at $z = 0.092$.\\

\noindent
\textbf{G4Jy 665} (GLEAM J073104$-$523710; Figure \ref{fig:g4jy665}): The radio morphology of G4Jy 665 is not quite clear in both SUMSS and TGSS contours, and as a result, this radio source was assigned a `complex' morphology label, host flag `u' and confusion flag `0'. The MeerKAT image reveals two unrelated radio sources, both having a WAT morphology and a detected host galaxy in the mid-infrared. The host galaxy of the southern source (indicated with a white `+' in the overlay) is AllWISE J073104.92$-$523808.6 at z = 0.09035, and the host galaxy of the northern source is AllWISE J073103.39$-$523546.8 at $z = 0.07690$. Given that the MeerKAT image reveals two unrelated sources with integrated flux densities of 0.836 Jy (southern source) and 0.314 Jy (northern source) at 1.3 GHz, this implies the integrated flux density at 151 MHz (5.12469 Jy) of G4Jy 665 is the sum of the integrated flux densities of the two sources. With this in mind, G4Jy 665 should be re-inspected in terms of the integrated flux density at 151 MHz, as this parameter is the G4Jy Sample's defining criterion ($S_{151~MHz} > 4$~Jy). \\

\noindent
\textbf{G4Jy 693} (GLEAM J081630$-$703925; Figure \ref{fig:g4jy693}): WAT radio morphology is evident in the MeerKAT contours, including detection of the radio core. An optical identification provided by \citet{1992ApJS...80..137J} does not coincide with the radio core detected in the MeerKAT image. The corresponding host galaxy is the mid-infrared AllWISE J081611.74-703945.3 at redshift z = 0.033. \\

\noindent
\textbf{G4Jy 1067} (GLEAM J132606$-$272641 and GLEAM J132616$-$272632, Figure \ref{fig:g4jy1067}): The radio contours from TGSS, NVSS and MeerKAT indicate WAT radio morphology, but the radio core is only detected in the MeerKAT image. The mid-infrared source that coincides with the radio core is AllWISE J132610.59$-$272538.6 at redshift $z = 0.044$. This radio source, known as PKS B1323$-$271, is in the cluster Abell 1736.

\begin{figure*}
\begin{multicols}{2}
\begin{subfigure}{\columnwidth}
\centering
\includegraphics[width=0.85\linewidth]{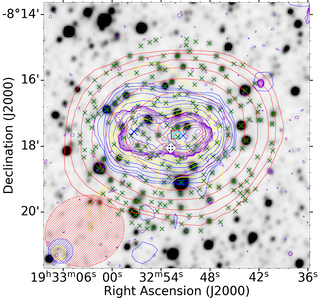}\par\caption{G4Jy 1554}\label{fig:g4jy1554}
\end{subfigure}

\begin{subfigure}{\columnwidth}
\centering
\includegraphics[width=1.\linewidth]{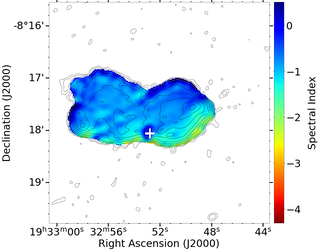}\par\caption{Spectral index map}\label{fig:index1554}
\end{subfigure}
\end{multicols}
\caption{Top panel: Overlay of G4Jy 1554. The datasets, symbols, beams and contours are the same as those described in Figure \ref{fig:g4jy40}. Bottom panel: The spectral index map with MeerKAT contours overlaid.}\label{fig:overlay_index_1554}
\end{figure*} 

\subsubsection{New identification via the spectral index map}
\textbf{G4Jy 1554} (GLEAM J193252$-$081739): \cite{2020PASA...37...17W, 2020PASA...37...18W} classified this radio source as `double' morphology based on NVSS and TGSS contours. `Complex' morphology is evident in the MeerKAT contours (Figure \ref{fig:g4jy1554}) with a slight indication of a radio core. The spectral index map [see Section \ref{subsubsec:meerkat}] (Figure \ref{fig:index1554}) indicates a radio core. We identify AllWISE J193252.79$-$081803.4 ($z = 0.101$) as the host galaxy.

\subsubsection{Giant radio galaxies (GRGs)}
\begin{figure}
\begin{subfigure}{\columnwidth}
\centering
\includegraphics[width=0.9\linewidth]{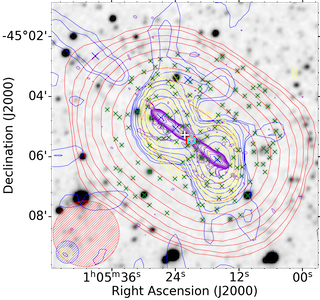}\caption{G4Jy 120}\label{fig:g4jy120}
\end{subfigure}

\begin{subfigure}{\columnwidth}
\centering
\includegraphics[width=0.9\linewidth]{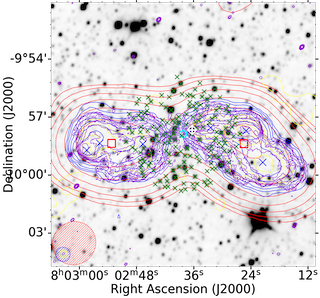}\caption{G4Jy 680}\label{fig:g4jy680}
\end{subfigure}
\caption{Radio galaxies with linear size above 0.7 Mpc in the MeerKAT-2019 subset. The datasets, symbols, beams and contours are the same as those described in Figure \ref{fig:g4jy40}.} \label{fig:GRGs}
\end{figure}

GRGs are a subclass of radio galaxies with a projected linear size > 0.7 Mpc. Their incredible Mpc-scale sizes make them one of the largest known objects in the universe. We find three sources (Figures \ref{fig:g4jy_1741} and \ref{fig:GRGs}) with linear sizes above 0.7 Mpc in the MeerKAT-2019 subset. \\

\noindent
\textbf{G4Jy 120} (GLEAM J010521$-$450527; Figure \ref{fig:g4jy120}): `Double' morphology is evident in the MeerKAT contours. We identify the host galaxy as AllWISE J010522.21$-$450517.2 at $z = 0.7$. This redshift value and the angular extent of $139.717~$arcsec correspond to the linear size of 1.004 Mpc.\\

\noindent
\textbf{G4Jy 680} (GLEAM J080225$-$095823 and GLEAM J080253$-$095822; Figure \ref{fig:g4jy680}): This radio source has several AllWISE positions that could be likely the host galaxy \citep{2020PASA...37...17W}. The MeerKAT image reveals the radio core which coincide with AllWISE J080236.28$-$095739.9, detected in 6dFGS as g0802363$-$095740. This identification is consistent with the optical identification of \citet{schilizzi1975observations}. The redshift of 0.070 and the angular extent of 9.13 arcmin at 1.4 GHz correspond to the linear size of $0.732~\mathrm{Mpc}$.\\

\subsubsection{Possible GRG}
\begin{figure}
\centering
\includegraphics[width=0.9\linewidth]{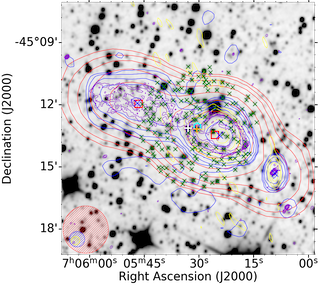}\caption{G4Jy 641: A possible GRG in the MeerKAT-2019 subset. The orange `+' sign indicate the optical identification provided by \citet{1992ApJS...80..137J}. The datasets, symbols, beams and contours are the same as those described in Figure \ref{fig:g4jy40}.}\label{fig:g4jy641}
\end{figure}

\textbf{G4Jy 641} (GLEAM J070525$-$451328 and GLEAM J070546$-$451158; PKS B0703$-$451): A possible GRG with radio emission that spans $\sim$8-arcmin from lobe to lobe in the 1.3 GHz image from MeerKAT (Figure \ref{fig:g4jy641}). `Triple' morphology is evident in the MeerKAT contours, and there is a detection of a hotspot in the southwestern lobe. However, there is an AllWISE position at the center of this hotspot emission, which makes it likely unrelated to G4Jy 641 An optical identification provided by \cite{1992ApJS...80..137J} (marked with the orange `+' sign in the overlay) appears to be incorrect. The corresponding host galaxy that coincides with the radio core is AllWISE J070532.94$-$451308.8. We find no redshift information in the literature and, therefore, cannot provide the linear size of this radio galaxy. 

\subsubsection{Quasi-stellar radio source}
\begin{figure}
\centering
\includegraphics[width=0.9\linewidth]{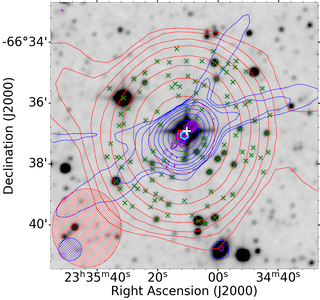}\caption{An overlay of a quasi stellar object (G4Jy 1843). The datasets, symbols, beams and contours are the same as those described in Figure \ref{fig:g4jy40}.}\label{fig:g4jy1843}
\end{figure}

\textbf{G4Jy 1843} (GLEAM J233511$-$663702; PKS B2332$-$62, Figure \ref{fig:g4jy1843}): This radio source has a `complex' morphology (likely artefacts) evident in the SUMSS contours, and there is no coverage in TGSS. The diffraction spikes evident in the mid-infrared image, and WISE colours, suggest that AllWISE J233510.30$-$663655.7 (close to the centroid position) is a star. However, the `double' morphology evident in the MeerKAT contours strongly suggest that this AllWISE source is likely the host galaxy of the radio emission (cf. the quasar 3C~273, which appears in the G4Jy Sample as G4Jy 1003; \citealt{2020PASA...37...17W}). We, therefore, regard AllWISE J233510.30$-$663655.7 as the host galaxy of this radio emission. 

\subsection{Unidentified sources}\label{sec:unidentifiedsources}
The overlays for MeerKAT-2019 subset sources discussed in this section are available as online supplementary material.
\subsubsection{Amorphous morphology}
\textbf{G4Jy 77} (GLEAM J004130$-$092221) is B0038$-$096 in the cluster Abell 85. This radio source is interpreted as a radio halo in the literature \citep{1998MNRAS.296L..23B}. The unusual morphology in the NVSS and TGSS contours is also evident in the MeerKAT image. We find that this radio source has the steepest spectral index ($\alpha^{1300~\rm{MHz}}_{151~\rm{MHz}} = -2.604$) in the MeerKAT-2019 subset, followed by G4Jy 1117 ($\alpha^{1300~\rm{MHz}}_{151~\rm{MHz}} = -2.018$). \\

\noindent
\textbf{G4Jy 513} (GLEAM J045826$-$300717, PKS B0456$-$30): This radio source in cluster A3297 has a dense field of mid-infrared AllWISE sources. \cite{1992ApJS...80..137J} interpreted the radio morphology as a cluster halo. The optical identification they provided coincide with the mid-infrared AllWISE source marked with the letter `A' in our overlay. This galaxy is at z = 0.131, which is consistent with the redshift of the cluster. We are unable to provide/confirm whether the provided optical identification by \cite{1992ApJS...80..137J} is the host galaxy for this radio emission due to the complex, amorphous morphology evident even in the higher-resolution image from MeerKAT. There is no detection of the radio core in neither the MeerKAT image nor the spectral index map.\\

\noindent
\textbf{G4Jy 700} (GLEAM J082231$+$055626): A radio source with diffuse emission known as 3C 198 in the 3C catalogue. \citet{wyndham1966optical} provided an optical identification corresponding to AllWISE J082231.95$+$055706.8 marked with the letter `A' in our overlay. However, there is no core detection in the higher-resolution image from MeerKAT nor in the spectral index map. The lack of core detection could be that the MeerKAT images are generated from 5-minute snapshot observations, and a longer integration time may be needed. The spectral index map reveals old emission in the inner regions, indicating that there is no ongoing ejection of electrons. This is the typical distribution of spectral indices in radio galaxies. Often the plasma near the centre is a result of backflow from the jets and is therefore older with a steeper spectral index.

\subsubsection{Unrelated sources}\label{subsec:unrelated}
Radio sources whose radio emission is blended together by NVSS/SUMSS (45 arcsec resolution) and TGSS (25 arcsec resolution) but MeerKAT images ($\sim$7 arcsec resolution) reveal two resolved sources. We find six sources (G4Jy 14, G4Jy 456, G4Jy 665, G4Jy 671, G4Jy 1491 and G4Jy 1815) in the MeerKAT-2019 subset where MeerKAT images (VLASS image for G4Jy 456) clearly reveal two unrelated sources. For G4Jy 671 and G4Jy 1491, the morphology in both NVSS/SUMSS and TGSS was uncertain, and the sources were therefore assigned `complex' morphology \citep{2020PASA...37...17W, 2020PASA...37...18W}. Given that the MeerKAT contours reveal two resolved, unrelated sources, we do not provide the properties (i.e, $\alpha^{1300~\rm{MHz}}_{151~\rm{MHz}}$, $L_{1300~\rm{MHz}}$ and linear size) of these sources since the integrated flux density at 151 MHz for these sources is the sum of the two resolved, unrelated sources which were blended by NVSS/SUMSS and TGSS. \\

\noindent
\textbf{G4Jy 456}: The extended, `double' morphology is evident in the NVSS and TGSS contours. To resolve the ambiguity of the host galaxy for this source, we obtained a higher-resolution radio image from archival VLASS (VLA Sky Survey [3 GHz and 2.5 arcsec resolution]; \citealt{2020PASP..132c5001L}) as the MeerKAT visibilities for this particular source, and G4Jy 453 are entirely flagged. The VLASS contours reveals two unrelated sources with `single` and `double' morphologies, respectively. The radio source with `single' morphology has a faint mid-infrared host, while the mid-infrared counterpart of the `double' morphology source is AllWISE J042536.76$+$083217.7 at $z = 1.1$.\\

\noindent
\textbf{G4Jy 671}: In both SUMSS and TGSS contours, the radio morphology of G4Jy 671 is not clear. The radio morphology could either be a head-tail, core-jet or `double' morphology with asymmetric jets. Hence this radio source was given the morphology label `complex',  host flag `u' and confusion flag '0'. The 1.3 GHz radio map from MeerKAT shows two unrelated radio sources with triple (northern source, $S_{1.3~\text{GHz}} = 1.062~Jy$) and `single' (southern source; $S_{1.3~\text{GHz}} = 0.237~$Jy) morphology, both having a detected host galaxy in mid-infrared. The host galaxy of the triple morphology source is AllWISE J074146.64$-$523415.9, and AllWISE J074151.46$-$523524.4 is the host galaxy of the radio source with `single' morphology. Given that there are two unrelated sources resolved by MeerKAT, the confusion flag of G4Jy 671 needs to be updated to `1'. \\

\noindent
\textbf{G4Jy 1491} (GLEAM J183356$-$394023): Based on the SUMSS and TGSS contours, it was unclear whether this radio source has a head-tail, `double' or core-jet morphology \citep{2020PASA...37...18W,2020PASA...37...17W}, and as such, it was assigned a 'complex' morphology label. The MeerKAT image indicates that there are two unrelated sources having `single' and `triple' morphology, respectively. The radio source with `single' morphology towards the north has no detected host galaxy in the mid-infrared, while the host galaxy of the southern source with `triple' morphology is AllWISE J183359.36$-$394158.3. The integrated flux density of G4Jy 1491 at 151 MHz is 4.24799 Jy. Given that the MeerKAT image reveals two unrelated sources, this implies the integrated flux density at 151 MHz for G4Jy 1491 is the sum of the integrated flux densities of the two sources resolved by MeerKAT. With this in mind, the integrated flux density at 151 MHz of G4Jy 1491 needs to be recalculated and the confusion flag updated to `1'.

\section{Discussion}\label{sec:discussion}
The primary aim of this work is to confirm/determine the host galaxy of the radio emission of 140 G4Jy sources (referred to as the MeerKAT-2019 subset) via visual inspection of overlays. Identifying the host galaxy of the radio source is required if we are to merge low-frequency radio data with other datasets, allowing us to develop a comprehensive, multi-wavelength view of the many processes occurring within these radio sources. Of the 140 sources making up the MeerKAT-2019 subset, we have identified the host galaxy of 98 sources. The host galaxy of G4Jy 1554 was identified through the inspection of the spectral index map, as the host could not be resolved through the overlay. We, therefore, assigned host flag `i' to the sources with an identified host galaxy. Of the remaining 42 sources with no identified host galaxy; 23 are assigned host flag `u', 18 have a faint mid-infrared host and are assigned host flag `m', and one source (G4Jy 77) is assigned host flag `n' as this radio source is identified as a radio halo. These MeerKAT-2019 subset sources still have an ambiguous host galaxy, even with higher-resolution images from MeerKAT.

\subsection{Brightness-weighted centroid to host-galaxy separation}\label{subsec:offsets}
\begin{figure}
\centering
\includegraphics[width=0.9\columnwidth]{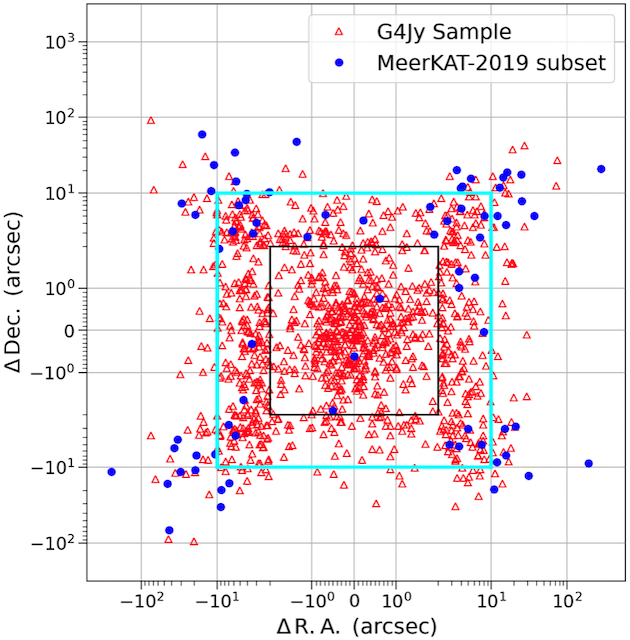}
\caption{The brightness-weighted centroid position to host galaxy position separation plot for the G4Jy Sample (red unfilled triangles) and MeerKAT-2019 subset (blue filled circles) sources with an identified host galaxy, ncmp\_GLEAM = `1' and confusion\_flag = `0'. The brightness-weighted centroid position is the weighted average of the NVSS/SUMSS components positions that are within the $3\sigma$ GLEAM contour of the G4Jy source. The ncmp\_GLEAM indicates the number of GLEAM components associated with the G4Jy source. We selected sources with ncmp\_GLEAM = `1' because sources with more than one GLEAM component have the same brightness-weighted centroid position. The confusion\_flag (either `0' or `1') indicates if the G4Jy source is affected by nearby unrelated emission detected above $6\sigma$ in NVSS/SUMSS and the position of the peak emission of this unrelated source is within the $3\sigma$ GLEAM contour level for the G4Jy source. We set the confusion\_flag to `0' to select sources that are not affected by nearby emission. The unusual square shape is because `symlog' (a symmetric log) is used for the x and y scale. This means that it is linear in the region $|\Delta RA| < 2"$ and $|\Delta Dec| < 2"$ (black square) and a log scale outside this region.} \label{fig:separation}
\end{figure}

Figure \ref{fig:separation} shows the separation between the brightness-weighted centroid position and host galaxy position for all the sources with an identified host galaxy, ncmp\_GLEAM = 1 and confusion\_flag = `0' in the MeerKAT-2019 subset and the G4Jy Sample. The brightness-weighted centroid position is the weighted average of the NVSS/SUMSS components positions, where the flux densities of these components are taken as weights. The NVSS/SUMSS components are associated with the G4Jy sources if they are within the $3\sigma$ GLEAM contour. The brightness-weighted centroid position is subject to error, if, for instance, the G4Jy source has two components in NVSS/SUMSS, and one component is two times brighter than the other component. Then the brightness-weighted centroid position will be shifted to the brighter component. This centroid position will then have a significant offset from the radio core and the host galaxy position. The median offset in R.A. is $0.108~\mathrm{arcsec}$, and the median offset in Dec. is $2.322~\mathrm{arcsec}$ for the MeerKAT-2019 subset. The median offset in R.A. and Dec. for the G4Jy Sample is $-0.144~\mathrm{arcsec}$ and $-0.040~\mathrm{arcsec}$, respectively. \\

\noindent
Overall, 1,215 sources in the G4Jy Sample have an identified host galaxy (sources with host flag `i'), ncmp\_GLEAM = `1' and confusion\_flag = `0', while the MeerKAT-2019 subset have 72 sources satisfying these criteria. About 10\% of the sources in the G4Jy Sample have a larger positional offset ($|\Delta\rm{R.A.}| > 10"$, $|\Delta\rm{Dec.}| > 10"$; sources outside the square cyan region), while the majority of the sources (57\%) in the MeerKAT-2019 subset populate this outer region (see Table~\ref{table:offsets}). We expect to have this larger positional offset for the MeerKAT-2019 subset because this subset is biased towards G4Jy sources with `complex' morphologies (i.e., head-tail, WAT and morphology that is not compact, nor typical symmetric double lobed) evident in NVSS/SUMSS and TGSS. For instance, sources with head-tail (Figure \ref{fig:head-tails}) and WAT (Figure \ref{fig:wat}) morphologies have their centroid position at the centre of the emission while their host galaxy is at the apex of the emission.

\begin{table*}
    \centering
    \begin{tabular}{c|c|c|c}
    \hline
      & Within the black outline & Between the black and cyan outlines & Outside of the cyan outline \\
     & (Approximately $|\Delta\rm{R.A.}| < 2''$, & (Approximately $2'' < |\Delta\rm{R.A.}| < 10''$, & (Approximately $|\Delta\rm{R.A.}| > 10''$,\\
     & $|\Delta\rm{Dec.}| < 2''$) &  $2'' < |\Delta\rm{Dec.}| < 10''$) & $|\Delta\rm{Dec.}| > 10''$) \\
    \hline
        G4Jy Sample & 41\% (496) & 49\% (597) & 10\% (122) \\
        MeerKAT-2019 subset & 4\% (3) & 39\% (28) & 57\% (41) \\
\hline
    \end{tabular}
    \caption{The fractions (and numbers) of G4Jy sources, within the full sample and MeerKAT-2019 subset, with different degrees of offset between the brightness-weighted centroid position and the host-galaxy position (see Subsection \ref{subsec:offsets}). `Black outline' and `cyan outline' are referring to the demarcations in Figure~\ref{fig:separation}. Note that these G4Jy sources have had the following criteria applied to them: host flag = `i', confusion flag = '0', and consisting of a single GLEAM component.}
    \label{table:offsets}
\end{table*}

\subsection{WISE colors}
\begin{figure}
    \centering
    \includegraphics[width=0.9\columnwidth]{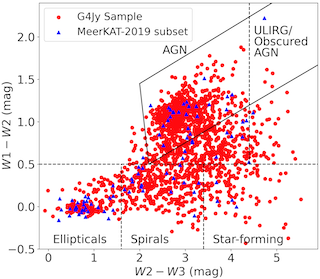}
    \caption{WISE colour-colour plot. W1, W2 and W3 correspond to wavelengths $3.4~\mu m$, $4.6~\mu m$ and $12~\mu m$.}
    \label{fig:wise}
\end{figure}

The WISE survey's four bands make it an excellent instrument for investigating galaxies' stellar structure and interstellar processes. The two shorter wavelength bands (W1 and W2) are used to map the stellar mass distribution in galaxies, whilst the longer wavelength bands (W3 and W4) are used to map warm dust emission and polycyclic aromatic hydrocarbon emission, which both trace the present star formation activity and AGN heating. Additionally, the WISE colours (W1 $-$ W2 and W2 $-$ W3) have been utilised to distinguish various astronomical objects (e.g., \citealt{Wright_2010, 2012MNRAS.426.3271M, 2016MNRAS.462.2631M}), which occupy distinct areas in the WISE colour-colour plot. The WISE colour-colour plot (Figure \ref{fig:wise}) is produced in this work to see where the G4Jy Sample and MeerKAT-2019 subset sources lie. Overall, 1,606 sources from the G4Jy Sample have host galaxy identified in AllWISE; however, 13 sources form part of the MeerKAT-2019 subset (see section \ref{sec:method} on the subset construction). The MeerKAT-2019 subset has 92 sources (including the 13) with AllWISE identification. Most sources in both the G4Jy Sample (63\%) and MeerKAT-2019 subset (42\%) occupy the AGN region (W1 $-$ W2 $>$ 0.5), with a handful of sources in the AGN wedge (solid black box) defined by \citet{2012MNRAS.426.3271M} for the G4Jy Sample. The G4Jy Sample is dominated by powerful AGN selected at low frequencies. There is a greater fraction of sources in the elliptical region for the MeerKAT-2019 subset (34\%) than in the G4Jy Sample (10\%). \\

\noindent
The elliptical and spiral regions are known to be dominated by low-excitation radio galaxies (LERGs) as these sources lack a dusty torus, while high-excitation radio galaxies (HERGs) dominate the region above W1 $-$ W2 $>$ 0.5. Additionally, LERGs are thought to be connected with FR-I, whereas HERGs are thought to be related to FRII. Several investigations, however, demonstrate that there is an overlap between HERG FR-Is and LERG FR-IIs (e.g., \citealt{2009MNRAS.396.1929H, 2012MNRAS.421.1569B, 2018MNRAS.480..358W, 2019MNRAS.488.2701M}).

\subsection{Radio properties}
\subsubsection{Spectral indices}
\begin{figure}
    \begin{subfigure}{\columnwidth}
    \centering
    \includegraphics[width=0.85\columnwidth]{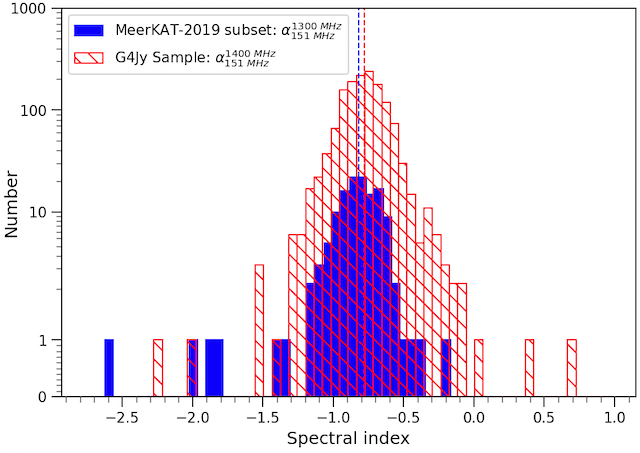}
    \end{subfigure}
    
    \begin{subfigure}{\columnwidth}
    \centering
    \includegraphics[width=0.87\columnwidth]{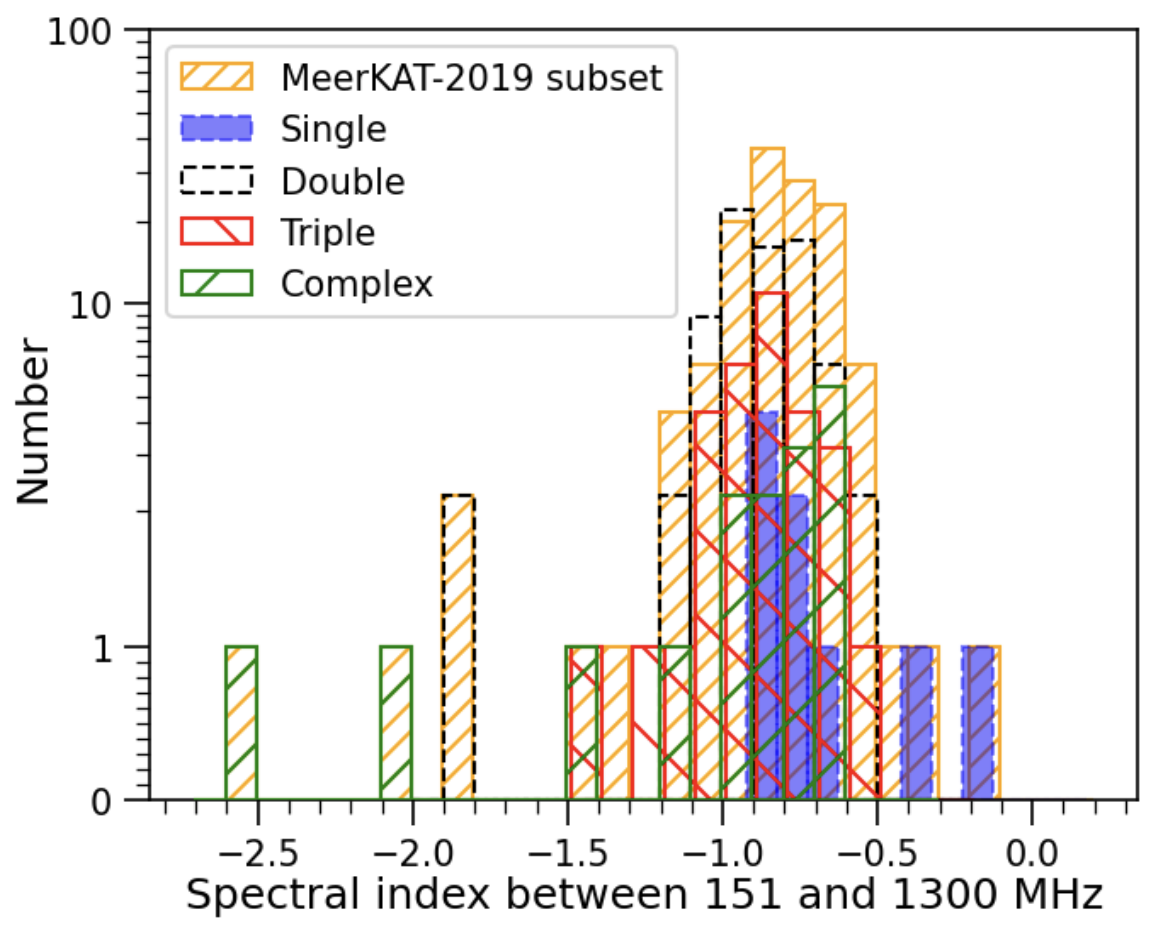}
    \end{subfigure}
\caption{Top panel: The distribution of the spectral index between 151 and 1300 MHz for the MeerKAT-2019 subset and the distribution of the spectral index between 151 and 1400 MHz for the G4Jy Sample. Bottom panel: The distribution of the spectral index between 151 and 1300 MHz for the full MeerKAT-2019 subset and sources with 'single', 'double', 'triple' and 'complex' morphologies.}\label{fig:sih}
\end{figure}

The two-point spectral index ($\alpha^{1300~\rm{MHz}}_{151~\rm{MHz}}$) between two radio frequencies (151 MHz and 1300 MHz) was calculated for 133\footnote{\label{note1}We have excluded sources where MeerKAT contours reveal two unrelated sources (see Section \ref{subsec:unrelated}).} sources in the MeerKAT-2019 subset, assuming $S \propto \nu^{\alpha}$, where $S$ is the integrated flux density at frequency $\nu$. The top panel of Figure \ref{fig:sih} shows the overall distribution of the two-point spectral index between 151 MHz and 1300 MHz for the MeerKAT-2019 subset compared with the distribution of the two-point spectral index between 151 MHz and 1400 MHz for the G4Jy Sample (obtained from the G4Jy catalogue). Both samples have a Gaussian-like distribution. The median spectral index of the G4Jy Sample and the MeerKAT-2019 subset are $-0.781$ and $-0.819$, respectively. One-hundred and twenty-seven (127) sources in the MeerKAT-2019 subset have a spectral index in the range $-1.2 < \alpha^{1300~\rm{MHz}}_{151~\rm{MHz}} < -0.5$, and the majority of these sources (74) have a `double' morphology (bottom panel of Figure \ref{fig:sih}). Six MeerKAT-2019 subset sources have an ultra-steep spectral index ($\alpha^{1300~\rm{MHz}}_{151~\rm{MHz}} < -1.2$), where G4Jy 77 is identified as a radio halo in the literature, G4Jy 1117 is likely a cluster relic, G4Jy 113 and G4Jy 1852 have head-tail morphology, and G4Jy 708 has a WAT morphology. The steep spectra of radio halos and cluster relics is due to synchrotron and Inverse Compton (IC) losses.

\subsubsection{Radio luminosity}
\begin{figure}
\begin{subfigure}{\columnwidth}
\includegraphics[width=0.9\columnwidth]{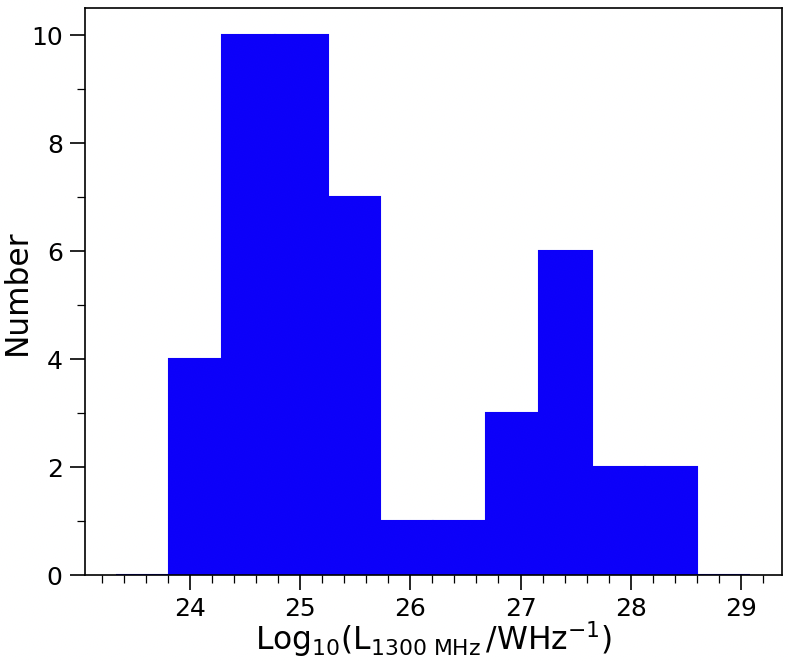}
\end{subfigure}
\begin{subfigure}{\columnwidth}
\includegraphics[width=0.9\columnwidth]{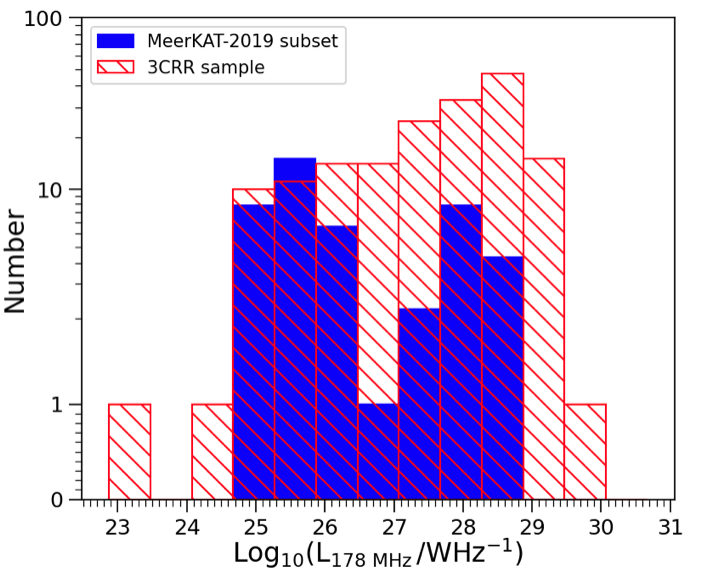}
\end{subfigure}
\caption{Top panel: The distribution of the 1.3 GHz radio luminosity for the MeerKAT-2019 subset. Bottom panel: For the purpose of comparison with the 3CRR sample, we calculated the integrated flux density at 178 MHz to obtain the radio luminosity at 178 MHz for the MeerKAT-2019 subset.}\label{fig:luminosity}
\end{figure}

\begin{figure}
    \centering
    \includegraphics[width=0.9\columnwidth]{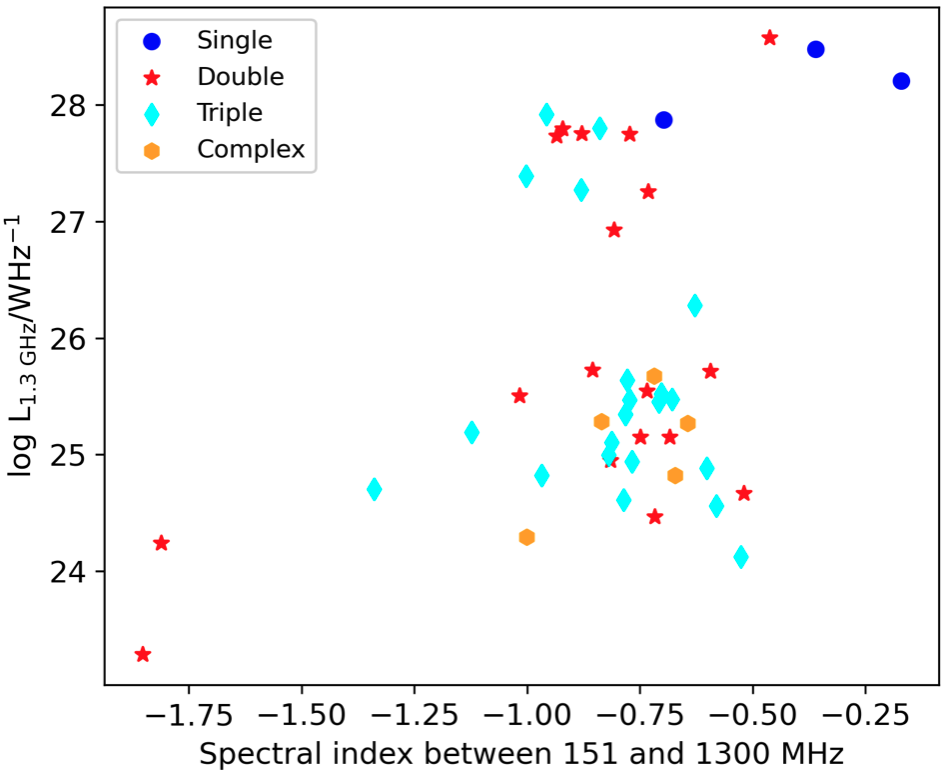}
    \caption{Plot of the two-point spectral index between 151 MHz and 1300 MHz vs. radio luminosity at 1300 MHz for the MeerKAT-2019 subset.} \label{fig:spl}
\end{figure}

After identifying the host galaxy of the MeerKAT-2019 subset sources, the next step was to obtain redshifts from existing catalogues/databases, so that we can calculate the radio luminosity at 1.3 GHz and the linear size of the MeerKAT-2019 subset. Of 98 radio sources with an identified host galaxy, 51 sources have redshift available from the literature. The 1.3 GHz radio luminosity ($L_{1.3~GHz}$) was calculated for 47\textsuperscript{\ref{note1}} of 51 sources using the equation: 
\begin{equation}
    L_{1.3~{\rm GHz}} = \frac{4\pi D_{L}^{2}S_{1.3~{\rm GHz}}}{(1+z)^{1+\alpha}}
\end{equation}

\noindent
where $D_L$ is the luminosity distance, $S_{1.3~GHz}$ is the integrated flux density at 1.3 GHz, $z$ is the redshift, and $\alpha$ is the spectral index between 151 MHz and 1300 MHz. The luminosity distance was calculated using Ned Wright's cosmology calculator\footnote{http://www.astro.ucla.edu/~wright/CosmoCalc.html}, assuming a flat Lambda CDM model ($H_0$ = 70 km/s/Mpc, $\Omega_m$ = 0.3 and $\Omega_{\Lambda} = 0.7$).  Figure \ref{fig:luminosity} shows the distribution of the 1.3 GHz radio luminosity of the sources in the MeerKAT-2019 subset where the redshift is available. Forty-three sources have radio luminosities in the order of $10^{24}$ W/Hz to $10^{27}$ W/Hz. These MeerKAT-2019 subset sources have a typical spectral index in the range $-1.2 < \alpha^{1300~\rm{MHz}}_{151~\rm{MHz}} < -0.5$. From Figure \ref{fig:spl}, we see that the MeerKAT-2019 subset sources with radio luminosities of the order $10^{28}$W/Hz have a flat spectral index ( $- 0.5 < \alpha^{1300~\rm{MHz}}_{151~\rm{MHz}} < 0.5$). The high luminosity of flat spectrum sources is due to the relativistic beaming effects. In contrast, sources with ultra-steep spectral index ($\alpha^{1300~\rm{MHz}}_{151~\rm{MHz}} \leq -1.2$) have radio luminosities of the order $10^{23}$ W/Hz to $10^{24}$ W/Hz. \\

\noindent
The bottom panel of Figure \ref{fig:luminosity} shows the distribution of the radio luminosities at 178 MHz ($L_{178~\rm{MHz}}$) of the MeerKAT-2019 subset compared with the radio luminosities 178 MHz ($L_{178~\rm{MHz}}$) of the 3CRR sample. For the radio luminosity at 178 MHz of the MeerKAT-2019 subset, we obtained (1) the integrated flux density at 181 MHz and (2) the G4Jy spectral index (calculated within the GLEAM band) from the G4Jy catalogue and calculated the integrated flux density at 178 MHz. The MeerKAT-2019 subset sources have typical radio luminosities of the $10^{25}$ W/Hz to $10^{29}$ W/Hz, while 3CRR sources have typical radio luminosities of the order $10^{24}$ to $10^{30}$ W/Hz.


\section{Conclusions}\label{sec:conclussion}
This study presents the analysis from studying 140 radio sources in the G4Jy Sample, referred to as the MeerKAT-2019 subset. Here, we summarise the work done in this study.
\begin{itemize}
    \item We constructed $\sim$10' by 10' overlays for all the sources in the MeerKAT-2019 subset. The overlays comprise radio data from GLEAM (200 MHz, $\sim$2-arcmin resolution), NVSS (1400 MHz, 45-arcsec resolution) /SUMSS (843 MHz, 45-arcsec resolution), TGSS (150 MHz, 25-arcsec resolution), and MeerKAT (1300 MHz, $\sim$7-arcsec resolution), overlaid onto mid-infrared WISE images ($3.4\mu m$).
    \item We visually inspected the overlays and classified 11 sources as `single' morphology, 78 as `double', 33 as `triple' and 16 as `complex' morphology radio sources. We note that most sources have radio morphology of typical symmetric lobes, while 10 sources have head-tail morphology, 14 have WAT morphology and 5 have X-, S-/Z-shaped morphology.
    \item We have identified the host galaxy of 98 sources in the MeerKAT-2019 subset, where the host galaxies were manually pinpointed through visual inspection. The 98 sources are assigned host flag `i', following \cite{2020PASA...37...17W, 2020PASA...37...18W} Of the remaining sources with no host galaxy identified, 23 are assigned host flag `u' as they have an ambiguous host even with higher resolution images from MeerKAT, G4Jy 77 is assigned host flag 'n' as it is identified as a radio halo in the literature, and 18 are assigned host flag `m' as they have a faint mid-infrared host. 
    \item Based on the brightness-weighted centroid position to host galaxy position separation plot, the MeerKAT-2019 subset have larger positional offset compared to the G4Jy Sample. The MeerKAT-2019 subset is biased towards the G4Jy sources with `complex' morphologies (including those with head-tail and WAT morphologies). We, therefore, expect to have these larger positional offsets. For instance, in head-tail sources, the host galaxy position will be near the apex of the radio emission instead of at the centre, where the brightness-weighted centroid position is.  
    \item We collected redshifts for 51 radio sources with an identified host galaxy in the MeerKAT-2019 subset and calculated their radio luminosity at 1.3 GHz. Radio sources with the steepest spectral index ($\alpha^{1300~\rm{MHz}}_{151~\rm{MHz}} < -1.2$) have low radio luminosities. These MeerKAT-2019 subset sources have typical head-tail morphology evident in MeerKAT contours. In contrast, sources with flat spectral index ($-0.5 < \alpha^{1300~\rm{MHz}}_{151~\rm{MHz}} < 0.5$)  have higher radio luminosities and redshifts above 1. The higher luminosity of compact, flat spectrum sources could be the result of relativistic beaming. \\
\end{itemize}

\noindent
This study has demonstrated the significance of a telescope's angular resolution and sensitivity for morphological classification and host-galaxy cross-identification of radio sources at low frequencies. The sensitivity and angular resolution ($\sim$7 arcsec) of MeerKAT has allowed us to resolve ambiguous host-galaxy identifications and enigmatic radio morphologies of the MeerKAT-2019 subset, evident in NVSS/SUMSS (45 arcsec) and/or TGSS (25 arcsec) images. 

These new identifications are crucial for gathering multi-wavelength data, with 42 of these MeerKAT-2019 subset sources having already been observed (2020-1-MLT-008; PI: White) with SALT. 

\section*{Acknowledgements}
The MeerKAT telescope is operated by the South African Radio Astronomy Observatory, which is a facility of the National Research Foundation, an agency of the Department of Science and Innovation. We acknowledge use of the Inter-University Institute for Data Intensive Astronomy (IDIA) data-intensive research cloud for data processing. IDIA is a South African university partnership involving the University of Cape Town, the University of Pretoria and the University of the Western Cape. The authors acknowledge the Centre for High-Performance Computing (CHPC), South Africa, for providing computational resources to this research project. We thank Chris Schollar, Masechaba Sydil Kupa, and Fernando Camilo for SARAO archive and DOI support. PKS and SVW acknowledge the financial assistance of the South African Radio Astronomy Observatory (SARAO; https://www.sarao.ac.za). IH also acknowledges support from SARAO, which is a facility of the National Research Foundation (NRF), an agency of the Department of Science and Innovation.

\section*{Data Availability}

For updates on the G4Jy Sample, please see https://github.com/svw26/G4Jy. The MeerKAT images and overlays (DOI: 10.48479/wyab-t838) can be accessed at \url{https://zenodo.org/communities/g4jy}, as well as via the SARAO archive \url{https://archive-gw-1.kat.ac.za/public/repository/10.48479/wyab-t838/index.html)}. The full Table \ref{tab:table5} and overlays for sources discussed in section \ref{sec:unidentifiedsources} are available as online supplementary material.



\bibliographystyle{mnras}
\bibliography{mnras_template} 



\bsp	
\label{lastpage}

\end{document}


\newgeometry{margin=0.33in,top=0.5in, bottom=0.5in}
\appendix
\onecolumn

\section{Properties of the MeerKAT-2019 subset}\label{apendix:tableresult}
This appendix provides the properties (Table \ref{tab:table5}) of the MeerKAT-2019 subset sources. 

\begin{center}
\captionof{table}{The full version of Table 3. Properties of sources in the MeerKAT-2019 subset. Columns 1-3 are; the G4Jy name, host flag, and the host galaxy name. Column 4 is the redshift, column 5 is the redshift reference, and column 6 is the spectral index ($\alpha$) between 151 and 1300 MHz. Columns 7-9 are the core intensity at 1300 MHz (SC13) in mJy/beam, integrated flux density at 1300 MHz (S13) in Jy, and integrated flux density at 1400 MHz (S14) in Jy. Columns 10-11 are the radio luminosities at 1300 MHz (L13) and 1400 MHz (L14) in W/Hz. Column 12 is the linear size (LS), and column 13 is the radio morphology label based on MeerKAT $\sim7-\text{arcsec}$ resolution images.} 
\label{tab:table5}
\small
 \begin{tabular}{cLcLLLccLccLc}
  \hline
G4Jy name & Host flag & Host name & z & z ref. & $\alpha$ & $SC13$ &  $S13$ & $S14$ & $L13$ & $L14$ & LS & Morphology \\ 
& & & & & & [mJy/beam] & [Jy] & [Jy] & [W/Hz] & [W/Hz] & [Mpc] & \\
  \hline
G4Jy 14 & i & WISEA J000707.12$+$053609.6 & 0.216 & (1) &  & $98.166$ & $0.441 \pm 0.085$ &  &  &  &  & Double \\
G4Jy 40 & i & WISEA J002107.53$-$191005.4 & 0.096 & (2) & $-0.707$ & $0.230$ & $1.261 \pm 0.140$ &  & $2.85\times 10^{25}$ &  &  & Triple \\
G4Jy 47 & i & WISEA J002531.45$-$330246.2 & 0.05 & (2) & $-0.815$ & $0.424$ & $1.527 \pm 0.162$ & 1.494 & $8.95\times 10^{24}$ & $8.75\times 10^{24}$ & 0.091  & Double\\
G4Jy 77 & u &  &  &  & $-2.604$ &  & $0.046 \pm 0.027$ & 0.035 &  &  &  & Complex \\
G4Jy 84 & i & WISEA J004605.01$-$633319.2 & 0.075 & (2) & $-0.82$ & $0.269$ & $0.731 \pm 0.100$ &  & $9.94\times 10^{24}$ &  & 0.215  & Triple \\
G4Jy 95 & u &  &  &  & $-0.924$ &  & $0.714 \pm 0.090$ & 0.668 &  &  &  & Double \\
G4Jy 113 & i & WISEA J010241.76$-$215254.2 & 0.057 & (2) & $-1.811$ & $2.172$ & $0.215 \pm 0.060$ & 0.199 & $1.75\times 10^{24}$ & $1.61\times 10^{24}$ & 0.132  & Double\\
G4Jy 120 & i & WISEA J010522.21$-$450517.2 & 0.71 & (3) & $-0.922$ & $3.154$ & $2.914 \pm 0.221$ &  & $6.28\times 10^{27}$ &  & 1.004  & Double \\
G4Jy 129 & i & WISEA J011141.93$-$685937.5 &  &  & $-0.812$ & $0.274$ & $2.027 \pm 0.158$ &  &  &  &  & Triple \\
G4Jy 284 & i & WISEA J023926.84$-$112752.6 &  &  & $-0.884$ & $4.072$ & $0.656 \pm 0.123$ & 0.592 &  &  &  & Double \\
G4Jy 285 & i & WISEA J024106.17$+$084416.9 & 0.021 & (4) & $-0.526$ & $-0.084$ & $1.346 \pm 0.160$ & 1.468 & $1.33\times 10^{24}$ & $1.45\times 10^{24}$ & 0.246  & Triple\\
G4Jy 311 & i & WISEA J025552.77$-$202749.0 &  &  & $-0.798$ & $0.919$ & $1.182 \pm 0.174$ & 1.071 &  &  &  & Double \\
G4Jy 318 & i & WISEA J030122.39$-$250445.1 &  &  & $-0.742$ & $1.022$ & $0.833 \pm 0.148$ & 0.77 &  &  &  & Triple \\
G4Jy 328 & i & WISEA J030919.39$-$103844.5 &  &  & $-0.587$ & $101.319$ & $1.192 \pm 0.168$ & 1.131 &  &  &  & Double \\
G4Jy 333 & i & WISEA J031154.01$-$313010.6 & 0.25 & (5) & $-0.628$ & $-0.037$ & $1.1 \pm 0.171$ & 1.052 & $1.92\times 10^{26}$ & $1.84\times 10^{26}$ & 0.416  & Triple\\
G4Jy 338 & u &  &  &  & $-1.007$ &  & $0.463 \pm 0.110$ & 0.484 &  &  &  & Double \\
G4Jy 343 & i & WISEA J031744.12$-$550628.2 & 0.806 & (6) & $-0.88$ & $15.237$ & $0.656 \pm 0.122$ &  & $1.88\times 10^{27}$ &  & 0.285  & Triple \\
G4Jy 350 & i & WISEA J032259.32$-$881600.4 &  &  & $-0.857$ & $12.729$ & $1.524 \pm 0.121$ &  &  &  &  & Double \\
G4Jy 365 & i & WISEA J033145.83$-$651754.2 &  &  & $-0.694$ & $27.093$ & $1.323 \pm 0.161$ &  &  &  &  & Double \\
G4Jy 372 & i & WISEA J033801.08$-$572344.8 &  &  & $-1.02$ & $153.245$ & $0.748 \pm 0.127$ &  &  &  &  & Double \\
G4Jy 375 & m &  &  &  & $-1.08$ &  & $0.396 \pm 0.088$ &  &  &  &  & Double \\
G4Jy 376 & i & WISEA J034046.51$-$340845.1 &  &  & $-0.894$ & $0.080$ & $0.669 \pm 0.129$ & 0.617 &  &  &  & Triple \\
G4Jy 413 & u &  &  &  & $-0.839$ &  & $0.766 \pm 0.143$ & 0.728 &  &  &  & Double \\
G4Jy 441 & u &  &  &  & $-0.64$ &  & $1.498 \pm 0.156$ & 1.36 &  &  &  & Double \\
G4Jy 453 & u &  &  &  &  &  &  &  &  &  &  &  \\
G4Jy 456 & i & WISEA J042536.76$+$083217.7 & 1.1 & (6) &  &  &  &  &  &  &  &  \\
G4Jy 472 & i & WISEA J043309.35$-$534346.4 &  &  & $-0.951$ & $2.222$ & $0.613 \pm 0.118$ &  &  &  &  & Triple \\
G4Jy 478 & i & WISEA J043539.49$-$494051.8 &  &  & $-0.839$ & $1.836$ & $0.777 \pm 0.138$ &  &  &  &  & Double \\
G4Jy 488 & i & WISEA J044111.01$+$251839.6 &  &  & $-0.922$ & $17.610$ & $0.595 \pm 0.087$ & 0.716 &  &  &  & Double \\
G4Jy 513 & u &  &  &  & $-0.692$ &  & $2.85 \pm 0.279$ & 2.779 &  &  &  & Complex \\
G4Jy 527 & m &  &  &  & $-0.64$ &  & $1.215 \pm 0.158$ & 1.119 &  &  &  & Double \\
G4Jy 526 & i & WISEA J051101.54$-$331600.8 &  &  & $-0.726$ & $8.790$ & $1.09 \pm 0.175$ & 0.998 &  &  &  & Double \\
G4Jy 528 & i & WISEA J051104.82$-$131730.3 & 0.043 & (2) & $-0.786$ & $9.890$ & $0.959 \pm 0.147$ & 0.957 & $4.11\times 10^{24}$ & $4.1\times 10^{24}$ & 0.086  & Triple\\
G4Jy 530 & i & WISEA J051247.41$-$482416.5 & 0.306 & (6) & $-0.807$ & $-0.446$ & $2.951 \pm 0.267$ &  & $8.46\times 10^{26}$ &  & 0.496  & Double \\
G4Jy 545 & i & WISEA J052615.97$+$114036.3 &  &  & $-0.864$ & $0.233$ & $1.014 \pm 0.126$ & 1.054 &  &  &  & Double \\
G4Jy 548 & i & WISEA J052857.49$+$151905.3 &  &  & $-0.78$ & $-0.026$ & $1.317 \pm 0.138$ & 1.208 &  &  &  & Triple \\
G4Jy 559 & u &  &  &  & $-0.722$ &  & $1.561 \pm 0.183$ &  &  &  &  & Double \\
G4Jy 566 & m &  &  &  & $-0.977$ &  & $0.506 \pm 0.082$ &  &  &  &  & Double \\
G4Jy 568 & i & WISEA J054016.90$-$330929.9 & 1.5 & (6) & $-0.957$ & $0.125$ & $0.609 \pm 0.122$ & 0.574 & $8.34\times 10^{27}$ & $7.82\times 10^{27}$ & 0.436  & Triple\\
G4Jy 570 & i & WISEA J054050.82$-$614237.2 & 0.081 & (3) & $-0.748$ & $7.425$ & $0.89 \pm 0.137$ &  & $1.41\times 10^{25}$ &  & 0.119  & Double \\
G4Jy 573 & i & WISEA J054515.99$-$315859.1 &  &  & $-0.93$ & $37.801$ & $0.598 \pm 0.120$ & 0.54 &  &  &  & Double \\
G4Jy 575 & i & WISEA J054617.91$-$172551.8 &  &  & $-0.728$ & $377.901$ & $1.137 \pm 0.159$ & 1.045 &  &  &  & Double \\
G4Jy 587 & u &  &  &  & $-0.768$ &  & $1.544 \pm 0.178$ &  &  &  &  & Complex \\
G4Jy 601 & i & WISEA J061637.48$-$142049.2 &  &  & $-0.734$ & $45.207$ & $1.117 \pm 0.157$ & 0.991 &  &  &  & Double \\
G4Jy 611 & i & WISEA J062620.46$-$534135.1 & 0.055 & (2) & $-0.855$ & $28.387$ & $7.43 \pm 0.420$ &  & $5.31\times 10^{25}$ &  & 0.064  & Double \\
G4Jy 623 & i & WISEA J063957.52$-$274542.5 &  &  & $-0.74$ & $0.641$ & $1.246 \pm 0.188$ & 1.182 &  &  &  & Double \\
G4Jy 637 & i & WISEA J070129.05$+$231325.6 & 0.092 & (4) & $-0.782$ & $43.508$ & $1.064 \pm 0.122$ & 0.958 & $2.22\times 10^{25}$ & $2\times 10^{25}$ & 0.176  & Triple\\
G4Jy 641 & i & WISEA J070532.94$-$451308.8 &  &  & $-0.876$ & $1.760$ & $0.835 \pm 0.147$ &  &  &  &  & Triple \\
G4Jy 643 & u &  &  &  & $-0.626$ &  & $1.058 \pm 0.165$ &  &  &  &  & Double \\
G4Jy 652 & i & WISEA J071818.14$-$580834.4 &  &  & $-0.819$ & $14.858$ & $0.701 \pm 0.130$ &  &  &  &  & Double \\
G4Jy 654 & m &  &  &  & $-0.845$ &  & $0.66 \pm 0.131$ & 0.58 &  &  &  & Single \\
G4Jy 665 & i & WISEA J073104.92$-$523808.6 & 0.09 & (2) &  & $2.970$ & $0.836 \pm 0.145$ &  &  &  &  & Triple \\
G4Jy 671 & i & WISEA J074146.64$-$523415.9 &  &  &  & $8.006$ & $1.062 \pm 0.162$ &  &  &  &  & Triple \\
		\hline
	\end{tabular}
\end{center}

\onecolumn

\begin{center}
 \label{tab:continued5.1}
 \small
 \begin{tabular}{cLcLLLccLccLc}
  \hline
G4Jy name & Host flag & Host name & z & z ref. & $\alpha$ & $SC13$ &  $S13$ & $S14$ & $L13$ & $L14$ & LS & Morphology \\ 
& & & & & & [mJy/beam] & [Jy] & [Jy] & [W/Hz] & [W/Hz] & [Mpc] & \\
\hline
G4Jy 672 & i & WISEA J074331.61$-$672625.5 & 1.51 & (2) & $-0.462$ & $485.179$ & $4.275 \pm 0.307$ &  & $3.77\times 10^{28}$ &  & 0.212  & Double \\
G4Jy 676 & m &  &  &  & $-0.847$ &  & $0.907 \pm 0.130$ &  &  &  &  & Single \\
G4Jy 677 & i & WISEA J075545.68$+$021029.0 &  &  & $-0.889$ & $44.878$ & $1.118 \pm 0.149$ & 1.008 &  &  &  & Double \\
G4Jy 680 & i & WISEA J080236.28$-$095739.9 & 0.07 & (2) & $-0.679$ & $0.253$ & $2.555 \pm 0.234$ & 2.174 & $2.98\times 10^{25}$ & $2.54\times 10^{25}$ & 0.732  & Triple\\
G4Jy 687 & i & WISEA J081158.15$-$052121.8 &  &  & $-0.868$ & $6.275$ & $0.787 \pm 0.122$ & 0.73 &  &  &  & Triple \\
G4Jy 693 & i & WISEA J081611.74$-$703945.3 & 0.033 & (2) & $-0.581$ & $0.328$ & $1.466 \pm 0.174$ &  & $3.63\times 10^{24}$ &  & 0.091  & Triple \\
G4Jy 700 & u &  &  &  & $-0.819$ &  & $2.409 \pm 0.218$ & 1.984 &  &  &  & Complex \\
G4Jy 708 & i & 2MASS J08284359$+$2437221 & 0.083 & (3) & $-1.339$ & $2.546$ & $0.29 \pm 0.099$ & 0.647 & $5.08\times 10^{24}$ & $1.1\times 10^{25}$ & 0.129  & Triple\\
G4Jy 717 & i & WISEA J083711.18$-$195156.6 & 1.032 & (6) & $-0.171$ & $2471.810$ & $5.161 \pm 0.353$ & 4.743 & $1.62\times 10^{28}$ & $1.52\times 10^{28}$ & 0.139  & Single\\
G4Jy 730 & i & WISEA J085300.25$-$204730.7 &  &  & $-0.81$ & $590.190$ & $2.476 \pm 0.241$ & 2.292 &  &  &  & Double \\
G4Jy 747 & i & WISEA J090147.26$-$255516.2 & 0.305 & (6) & $-0.732$ & $333.058$ & $6.489 \pm 0.395$ & 6.253 & $1.81\times 10^{27}$ & $1.74\times 10^{27}$ & 0.217  & Double\\
G4Jy 758 & i & WISEA J091237.88$-$641130.6 &  &  & $-0.864$ & $0.181$ & $0.626 \pm 0.116$ &  &  &  &  & Double \\
G4Jy 781 & u &  &  &  & $-0.915$ &  & $0.587 \pm 0.102$ & 0.54 &  &  &  & Double \\
G4Jy 869 & i & WISEA J104645.27$-$360138.5 &  &  & $-0.833$ & $26.957$ & $1.459 \pm 0.162$ & 1.32 &  &  &  & Double \\
G4Jy 901 & i & WISEA J111119.43$-$403051.9 &  &  & $-0.699$ & $253.768$ & $1.27 \pm 0.146$ &  &  &  &  & Double \\
G4Jy 917 & i & WISEA J112552.95$-$352340.3 & 0.033 & (2) & $-0.671$ & $3.340$ & $2.669 \pm 0.203$ & 2.606 & $6.63\times 10^{24}$ & $6.48\times 10^{24}$ & 0.068  & Complex\\
G4Jy 935 & u &  &  &  & $-0.69$ &  & $0.934 \pm 0.122$ &  &  &  &  & Complex \\
G4Jy 939 & u &  &  &  & $-0.673$ &  & $2.806 \pm 0.195$ & 2.571 &  &  &  & Double \\
G4Jy 984 & i & WISEA J121738.77$+$033948.2 & 0.078 & (1) & $-0.834$ & $44.599$ & $1.306 \pm 0.159$ & 1.116 & $1.93\times 10^{25}$ & $1.66\times 10^{25}$ & 0.037  & Complex\\
G4Jy 1067 & i & WISEA J132610.59$-$272538.6 & 0.044 & (2) & $-0.602$ & $2.064$ & $1.715 \pm 0.223$ & 1.595 & $7.65\times 10^{24}$ & $7.12\times 10^{24}$ & 0.148  & Triple\\
G4Jy 1073 & i & WISEA J133225.71$-$330809.9 & 0.049 & (4) & $-0.768$ & $1.474$ & $1.553 \pm 0.214$ & 1.423 & $8.71\times 10^{24}$ & $7.98\times 10^{24}$ & 0.062  & Triple\\
G4Jy 1094 & i & WISEA J134854.17$-$252724.5 & 0.126 & (2) & $-1.123$ & $0.137$ & $0.367 \pm 0.103$ & 0.329 & $1.55\times 10^{25}$ & $1.39\times 10^{25}$ & 0.175  & Triple\\
G4Jy 1097 & u &  &  &  & $-0.643$ &  & $1.076 \pm 0.167$ & 1.059 &  &  &  & Complex \\
G4Jy 1117 & u &  &  &  & $-2.018$ &  & $0.102 \pm 0.055$ & 0.154 &  &  &  & Complex \\
G4Jy 1190 & i & WISEA J144635.50$-$084605.3 & 0.071 & (3) & $-0.813$ & $8.772$ & $1.05 \pm 0.169$ & 0.956 & $1.27\times 10^{25}$ & $1.16\times 10^{25}$ & 0.092  & Triple\\
G4Jy 1201 & i & WISEA J145422.82$-$192509.2 &  &  & $-0.779$ & $90.105$ & $1.157 \pm 0.183$ & 1.1 &  &  &  & Double \\
G4Jy 1205 & i & WISEA J145509.61$-$365507.4 & 0.095 & (2) & $-0.773$ & $2.128$ & $1.319 \pm 0.197$ & 1.19 & $2.94\times 10^{25}$ & $2.65\times 10^{25}$ & 0.16  & Triple\\
G4Jy 1220 & u &  &  &  & $-1.414$ &  & $0.374 \pm 0.087$ &  &  &  &  & Complex \\
G4Jy 1247 & i & WISEA J152132.67$-$400104.0 &  &  & $-1.146$ & $1.752$ & $0.497 \pm 0.120$ &  &  &  &  & Triple \\
G4Jy 1260 & i & WISEA J152659.45$-$135100.0 & 1.687 & (6) & $-0.36$ & $2482.420$ & $2.991 \pm 0.288$ & 2.865 & $3.02\times 10^{28}$ & $2.92\times 10^{28}$ & 0.132  & Single\\
G4Jy 1262 & i & WISEA J153014.29$-$423151.7 & 0.5 & (3) & $-0.934$ & $4.687$ & $5.797 \pm 0.405$ &  & $5.42\times 10^{27}$ &  & 0.518  & Double \\
G4Jy 1289 & u &  &  &  & $-0.833$ &  & $2.13 \pm 0.241$ & 2.003 &  &  &  & Double \\
G4Jy 1297 & i & WISEA J160240.38$+$154521.1 & 0.037 & (4) & $-0.718$ & $0.205$ & $0.928 \pm 0.124$ & 0.929 & $2.92\times 10^{24}$ & $2.92\times 10^{24}$ & 0.066  & Double\\
G4Jy 1305 & m &  &  &  & $-0.875$ &  & $0.69 \pm 0.137$ & 0.618 &  &  &  & Single \\
G4Jy 1311 & i & WISEA J161338.86$-$771931.5 & 0.099 & (3) & $-0.702$ & $-0.106$ & $1.369 \pm 0.165$ &  & $3.3\times 10^{25}$ &  & 0.13  & Triple \\
G4Jy 1324 & i & WISEA J162035.89$-$710006.5 &  &  & $-0.899$ & $6.835$ & $0.975 \pm 0.147$ &  &  &  &  & Triple \\
G4Jy 1329 & m &  &  &  & $-0.928$ &  & $0.693 \pm 0.142$ & 0.593 &  &  &  & Double \\
G4Jy 1342 & i & WISEA J163135.04$-$750905.9 & 0.11 & (2) & $-1.017$ & $8.626$ & $1.024 \pm 0.145$ &  & $3.19\times 10^{25}$ &  & 0.137  & Double \\
G4Jy 1362 & m &  &  &  & $-0.897$ &  & $0.982 \pm 0.164$ & 0.898 &  &  &  & Double \\
G4Jy 1365 & i & WISEA J164604.85$-$222804.6 & 0.799 & (6) & $-0.773$ & $625.672$ & $2.15 \pm 0.243$ & 1.994 & $5.65\times 10^{27}$ & $5.27\times 10^{27}$ & 0.107  & Double\\
G4Jy 1371 & i & PanSTARRS 108382532433733198 &  &  & $-0.664$ & $1420.916$ & $1.442 \pm 0.179$ & 1.329 &  &  &  & Single \\
G4Jy 1377 & i & WISEA J165712.85$-$134909.5 & 0.124 & (3) & $-0.593$ & $8.625$ & $1.355 \pm 0.190$ & 1.311 & $5.2\times 10^{25}$ & $5.03\times 10^{25}$ & 0.146  & Double\\
G4Jy 1403 & i & WISEA J172040.93$-$011157.2 & 0.029 & (4) & $-1.001$ & $0.942$ & $1.019 \pm 0.149$ & 0.761 & $1.97\times 10^{24}$ & $1.47\times 10^{24}$ & 0.058  & Complex\\
G4Jy 1410 & i & WISEA J172437.79$-$024305.6 & 0.033 & (2) & $-0.519$ & $8.876$ & $1.877 \pm 0.206$ & 1.8538 & $4.64\times 10^{24}$ & $4.58\times 10^{24}$ & 0.043  & Double\\
G4Jy 1415 & i & WISEA J173122.25$+$210501.1 &  &  & $-0.844$ & $-1.416$ & $0.662 \pm 0.099$ & 0.653 &  &  &  & Triple \\
G4Jy 1429 & u &  &  &  & $-0.618$ &  & $2.162 \pm 0.241$ &  &  &  &  & Double \\
G4Jy 1431 & m &  &  &  & $-0.787$ &  & $1.027 \pm 0.142$ & 0.936 &  &  &  & Double \\
G4Jy 1438 & u &  &  &  & $-0.956$ &  & $0.766 \pm 0.102$ & 1.409 &  &  &  & Complex \\
G4Jy 1444 & m &  &  &  & $-0.722$ &  & $1.142 \pm 0.169$ &  &  &  &  & Double \\
G4Jy 1449 & m &  &  &  & $-0.9$ &  & $0.821 \pm 0.125$ & 0.717 &  &  &  & Double \\
G4Jy 1450 & i & WISEA J175824.93$+$082826.9 &  &  & $-0.762$ & $116.874$ & $0.835 \pm 0.126$ & 0.756 &  &  &  & Double \\
G4Jy 1452 & i & WISEA J175851.11$-$673829.0 &  &  & $-1.109$ & $-0.297$ & $0.637 \pm 0.124$ &  &  &  &  & Double \\
G4Jy 1455 & i & WISEA J180117.98$-$662301.8 & 0.93 & (3) & $-0.697$ & $285.146$ & $2.1 \pm 0.226$ &  & $7.51\times 10^{27}$ &  & 0.486  & Single \\
G4Jy 1464 & m &  &  &  & $-0.814$ &  & $1.038 \pm 0.172$ &  &  &  &  & Double \\
G4Jy 1468 & i & WISEA J181231.40$+$262916.6 &  &  & $-0.93$ & $40.300$ & $1.183 \pm 0.122$ & 1.115 &  &  &  & Double \\
G4Jy 1469 & m &  &  &  & $-1.006$ &  & $0.652 \pm 0.126$ &  &  &  &  & Double \\
G4Jy 1473 & u &  &  &  & $-1.132$ &  & $0.39 \pm 0.104$ &  &  &  &  & Complex \\
G4Jy 1474 & m &  &  &  & $-0.878$ &  & $2.796 \pm 0.289$ & 2.711 &  &  &  & Double \\
G4Jy 1491 & m &  &  &  &  &  & $1.153 \pm 0.186$ &  &  &  &  & Single \\
G4Jy 1493 & m &  &  &  & $-1.031$ &  & $0.482 \pm 0.077$ & 0.422 &  &  &  & Double \\
G4Jy 1494 & u &  &  &  & $-0.986$ &  & $0.81 \pm 0.142$ &  &  &  &  & Double \\
  \hline
 \end{tabular}
\end{center}

\begin{table}
 \label{tab:continued5.2}
 \begin{tabular}{cLcLLLccLccLc}
  \hline
G4Jy name & Host flag & Host name & z & z ref. & $\alpha$ & $SC13$ &  $S13$ & $S14$ & $L13$ & $L14$ & LS & Morphology \\ 
& & & & & & [mJy/beam] & [Jy] & [Jy] & [W/Hz] & [W/Hz] & [Mpc] & \\
  \hline
G4Jy 1501 & i & WISEA J184037.09$-$293353.9 &  &  & $-0.909$ & $13.469$ & $0.625 \pm 0.138$ & 0.745 &  &  &  & Double \\
G4Jy 1503 & i & WISEA J184214.94$-$351753.9 &  &  & $-0.944$ & $33.857$ & $0.636 \pm 0.139$ & 0.629 &  &  &  & Double \\
G4Jy 1518 & i & WISEA J191548.68$-$265257.4 & 0.226 & (3) &  & $-5.869$ & $1.608 \pm 0.219$ &  &  &  &  & Double \\
G4Jy 1523 & i & 2MASS J19175722$-$2439053 &  &  & $-0.754$ & $12.587$ & $0.853 \pm 0.159$ & 0.836 &  &  &  & Complex \\
G4Jy 1527 & i & WISEA J191947.64$-$525301.7 & 1.0 & (6) & $-1.002$ & $0.546$ & $0.47 \pm 0.115$ &  & $2.46\times 10^{27}$ &  & 0.388  & Triple \\
G4Jy 1534 & u &  &  &  & $-0.889$ &  & $0.635 \pm 0.137$ &  &  &  &  & Double \\
G4Jy 1537 & i & WISEA J192605.75$-$574016.4 & 0.061 & (2) & $-0.968$ & $0.749$ & $0.742 \pm 0.140$ &  & $6.62\times 10^{24}$ &  & 0.166  & Triple \\
G4Jy 1539 & m &  &  &  & $-0.678$ &  & $1.111 \pm 0.182$ & 0.848 &  &  &  & Single \\
G4Jy 1540 & i & PanSTARRS 80412918521248951 &  &  & $-0.957$ & $-0.058$ & $0.607 \pm 0.134$ & 0.603 &  &  &  & Double \\
G4Jy 1548 & i & 2MASS J19313531$-$1406009 &  &  & $-0.99$ & $5.554$ & $0.859 \pm 0.157$ & 1.226 &  &  &  & Double \\
G4Jy 1553 & u &  &  &  & $-0.803$ &  & $1.325 \pm 0.199$ & 1.262 &  &  &  & Double \\
G4Jy 1554 & i & WISEA J193252.79$-$081803.4 & 0.101 & (2) & $-0.718$ & $5.761$ & $1.867 \pm 0.227$ & 1.85 & $4.71\times 10^{25}$ & $4.65\times 10^{25}$ & 0.167  & Complex\\
G4Jy 1582 & i & WISEA J195230.53$-$011720.7 & 0.055 & (2) & $-0.684$ & $3.249$ & $2.007 \pm 0.224$ & 1.618 & $1.42\times 10^{25}$ & $1.15\times 10^{25}$ & 0.349  & Double\\
G4Jy 1585 & i & WISEA J195513.78$+$015409.3 &  &  & $-0.853$ & $55.181$ & $0.787 \pm 0.135$ & 1.013 &  &  &  & Double \\
G4Jy 1617 & i & WISEA J202338.06$+$170236.1 &  &  & $-0.649$ & $0.037$ & $1.221 \pm 0.142$ & 1.172 &  &  &  & Triple \\
G4Jy 1638 & i & WISEA J203444.74$-$354901.7 & 0.089 & (2) & $-0.735$ & $78.395$ & $1.811 \pm 0.229$ & 1.717 & $3.5\times 10^{25}$ & $3.32\times 10^{25}$ & 0.04  & Double\\
G4Jy 1663 & m &  &  &  & $-0.848$ &  & $0.655 \pm 0.136$ & 0.598 &  &  &  & Single \\
G4Jy 1697 & i & WISEA J212937.79$-$224824.4 &  &  & $-0.791$ & $95.595$ & $0.896 \pm 0.157$ & 0.833 &  &  &  & Double \\
G4Jy 1706 & i & WISEA J213516.47$-$173433.4 &  &  & $-0.687$ & $442.701$ & $1.132 \pm 0.174$ & 1.06 &  &  &  & Double \\
G4Jy 1740 & i & WISEA J215407.02$-$515012.8 &  &  & $-0.7$ & $37.574$ & $3.845 \pm 0.322$ &  &  &  &  & Double \\
G4Jy 1741 & i & WISEA J215422.96$-$455231.3 & 0.145 & (2) & $-0.778$ & $5.818$ & $0.793 \pm 0.148$ &  & $4.35\times 10^{25}$ &  & 0.882  & Triple \\
G4Jy 1784 & i & WISEA J223746.49$-$114538.7 & 1.0 & (6) & $-0.878$ & $113.874$ & $1.191 \pm 0.168$ & 1.634 & $5.72\times 10^{27}$ & $6.97\times 10^{27}$ & 0.197  & Double\\
G4Jy 1798 & i & 2MASS J22564194$-$4617351 & 0.08 & (3) & $-0.645$ & $104.001$ & $1.202 \pm 0.174$ &  & $1.85\times 10^{25}$ &  & 0.087  & Complex \\
G4Jy 1843 & i & WISEA J233510.30$-$663655.7 & & & $-0.69$ & $4.332$ & $2.472 \pm 0.227$ &  & &  & & Double \\
G4Jy 1852 & i & WISEA J234745.06$-$280826.2 & 0.029 & (2) & $-1.853$ & $2.041$ & $0.098 \pm 0.047$ & 0.085 & $1.94\times 10^{23}$ & $1.68\times 10^{23}$ & 0.068  & Double\\
G4Jy 1859 & i & WISEA J235706.20$-$022618.6 & 1.2779 &  & $-0.839$ & $-2.841$ & $0.751 \pm 0.116$ & 0.767 & $6.3\times 10^{27}$ & $6.24\times 10^{27}$ & 0.491  & Triple\\
  \hline
  \hline
\multicolumn{13}{c}{  } \\
\multicolumn{13}{c}{Redshift references: (1) Alam et al. (2015), (2) Jones et al. (2004), (3) NASA/IPAC Extragalactic Database (NED), (4) Huchra et al. (2012), (5) de Jong et al. (2017), and} \\
\multicolumn{2}{c}{(6) Flesch (2015)} & & & & & & & & & & \\

 \end{tabular}
\end{table}

\onecolumn
\section{Sources with amorphous morphology}\label{appendix:amorphous}
This appendix provides the overlays of MeerKAT-2019 subset sources with amorphous morphology (section 4.2.1). \\

\begin{center}
\centering
\includegraphics[width=0.54\columnwidth]{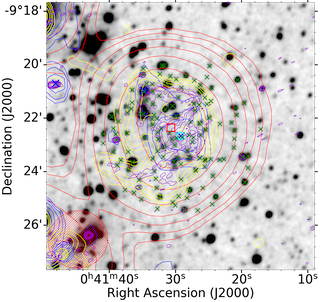} 
\par\bigskip
\centering
\includegraphics[width=0.56\columnwidth]{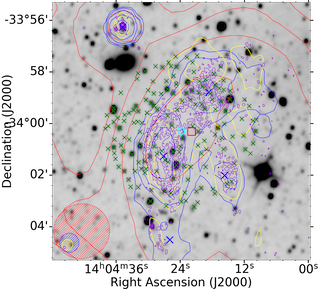} 

\captionof{figure}{Top panel: An overlay of G4Jy 77. This radio source is identified as a radio halo in the literature. Bottom panel: An overlay of G4Jy 1117 (likely a cluster relic). The radio contours are from GLEAM (170 - 231 MHz; red), NVSS (1.4 GHz; blue), TGSS (150 MHz; yellow), and MeerKAT (1.3 GHz; purple) overlaid on the inverted grayscale WISE (3.4 $\mu m$) image. Shown in the bottom left corner is the beam size for each survey; GLEAM (red), NVSS(blue), TGSS (yellow) and MeerKAT (purple). The cyan hexagon indicates the brightness-weighted centroid position, and the green crosses `x' signs are the AllWISE positions within 3 arcmin from the centroid position. The red square is the GLEAM component position, and blue crosses are NVSS/SUMSS components positions.}
\label{fig:g4jy77}
\end{center}

\begin{figure*}
\begin{subfigure}{\columnwidth}
\centering
\includegraphics[width=0.7\columnwidth]{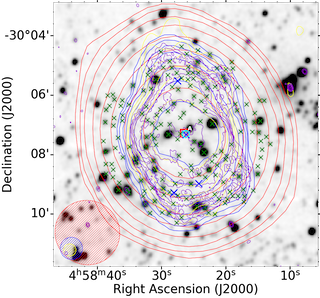}
\end{subfigure}

\begin{subfigure}{\columnwidth}
\centering
\includegraphics[width=0.7\columnwidth]{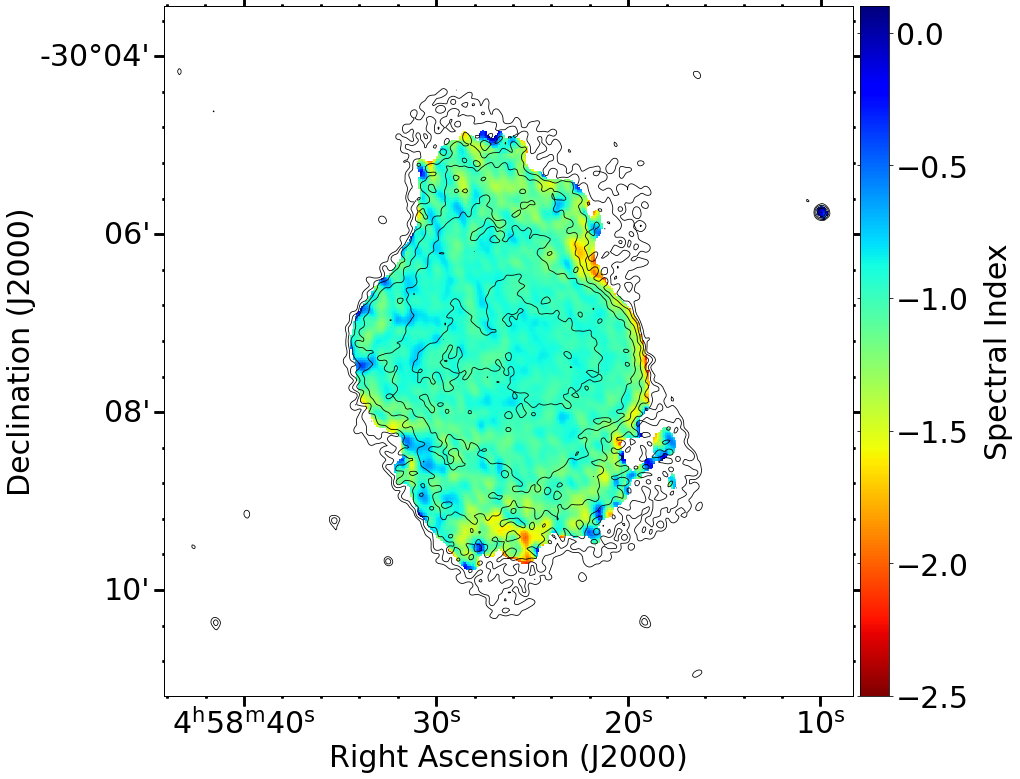}
\end{subfigure}

\caption{Top panel: An overlay of G4Jy 513. This radio source has `complex' morphology evident in all surveys. The optical identification of Jones \& McAdam (1992) is marked with the letter `A', which we are unable to confirm as there is no definitive core in the overlay. The datasets, symbols, beams and contours are the same as those described in Figure \ref{fig:g4jy77}. Bottom panel: The spectral index map of G4Jy 513 with MeerKAT contours overlaid. The spectrum is almost steep everywhere, making it difficult to locate the radio core.}
\end{figure*}

\begin{figure*}
\begin{subfigure}{\columnwidth}
\centering
\includegraphics[width=0.7\columnwidth]{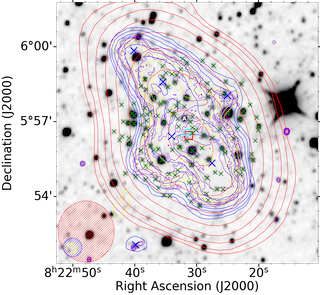}
\end{subfigure}

\begin{subfigure}{\columnwidth}
\centering
\includegraphics[width=0.7\columnwidth]{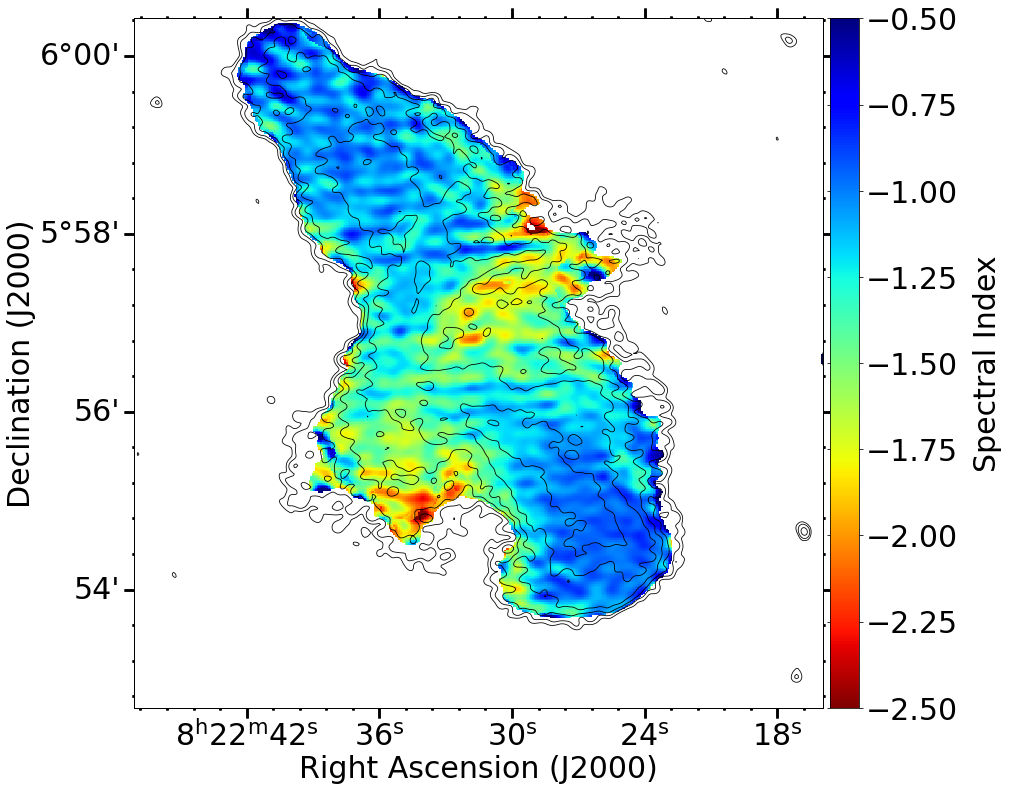}
\end{subfigure}

\caption{Top panel: An overlay of G4Jy 700. The letter `A' is the optical identification provided by Wyndham (1966), which we are unable to confirm. The datasets, symbols, beams and contours are the same as those described in Figure \ref{fig:g4jy77}. Bottom panel: The spectral index map of G4Jy 700 with MeerKAT contours overlaid. The steepening toward the central region is a possible indication of the  plasma backflow from the jets.}
\end{figure*}

\section{Overlays of unrelated sources} \label{appendix:unrelated}
This appendix provides the overlays of six MeerKAT-2019 subset sources where the MeerKAT contours (VLASS for G4Jy 456) reveals two unrelated sources (subsection 4.2.2).

\begin{figure*}
\begin{multicols}{2}
\begin{subfigure}{\columnwidth}
\centering
\includegraphics[width=0.8\columnwidth]{Figures2/Overlays/GLEAM_J000707+053607_NVSS_hres_12_sigma_v2.png}\par\caption{ G4Jy 14}\label{fig:g4jy14a}
\end{subfigure}

\begin{subfigure}{\columnwidth}
\centering
\includegraphics[width=0.8\columnwidth]{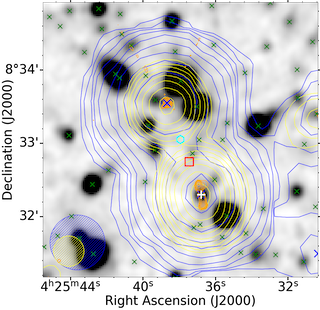}\par\caption{ G4Jy 456}\label{fig:g4jy456}
\end{subfigure}
\end{multicols}

\begin{multicols}{2}
\begin{subfigure}{\columnwidth}
\centering
\includegraphics[width=0.8\columnwidth]{Figures2/Overlays/GLEAM_J073104-523710_SUMSS_3_sigma_v2.png}\par\caption{ G4Jy 665}\label{fig:g4jy665a}
\end{subfigure}

\begin{subfigure}{\columnwidth}
\centering
\includegraphics[width=0.8\columnwidth]{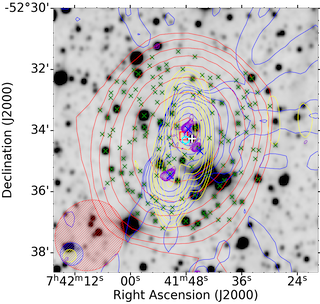}\par\caption{ G4Jy 671}\label{fig:g4jy671}
\end{subfigure}
\end{multicols}

\begin{multicols}{2}
\begin{subfigure}{\columnwidth}
\centering
\includegraphics[width=0.8\columnwidth]{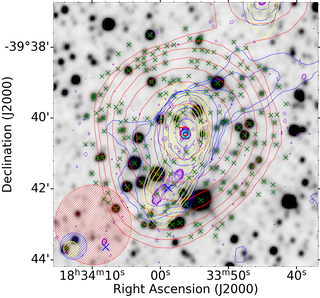}\par\caption{G4Jy 1491}\label{fig:g4jy1491}
\end{subfigure}

\begin{subfigure}{\columnwidth}
\centering
\includegraphics[width=0.8\columnwidth]{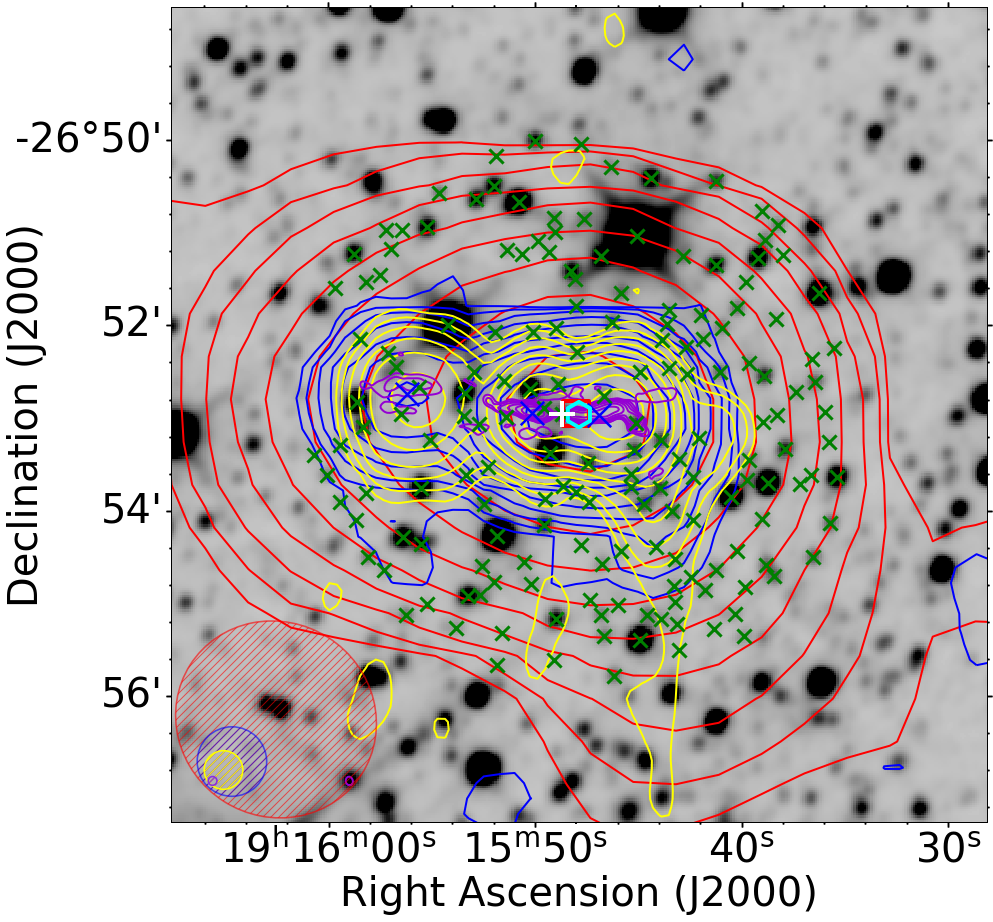}\par\caption{G4Jy 1518}\label{fig:g4jy1518}
\end{subfigure}
\end{multicols}

\caption{Radio sources in the MeerKAT-2019 subset whose radio emission is blended together by NVSS/SUMSS and TGSS. The VLASS contours (orange) in the overlay of G4Jy 456 shows two unrelated sources. The datasets, symbols, beams and contours are the same as those described in Figure \ref{fig:g4jy77}.}\label{fig:unrelateds}
\end{figure*}